# Deep Learning Inter-atomic Potential for Thermal and Phonon Behaviour of Silicon Carbide with Quantum Accuracy

Baoqin Fu[a], Yandong Sun[b], Linfeng Zhang[c], Han Wang[d], Ben Xu[b*]

[a]Key Laboratory of Radiation Physics and Technology of the Ministry of Education, Institute of Nuclear Science and Technology, Sichuan University, Chengdu 610064, People's Republic of China

[b]Graduate School, China Academy of Engineering Physics, Beijing 100193, People's Republic of China

[c]Program in Applied and Computational Mathematics, Princeton University, Princeton, NJ, USA

[d]Laboratory of Computational Physics, Institute of Applied Physics and Computational Mathematics, Beijing 100088, People's Republic of China

**Abstract**

Silicon carbide (SiC) is an essential material for next generation semiconductors and components for nuclear plants. It's applications are strongly dependent on its thermal conductivity, which is highly sensitive to microstructures. Molecular dynamics (MD) simulation is the most used methods to address thermal transportation mechanisms in devices or microstructures of nano-meters. However, the implementation of MD is limited in SiC because of lacking accurate inter-atomic potentials. In this work, using the Deep Potential (DP) methodology, we developed two inter-atomic potentials (DP-

[*]Corresponding author.

IAPs) for SiC based on two adaptively generated datasets within the density functional approximations at the local density and the generalized gradient levels. These two DP-IAPs manifest their speed with quantum accuracy in lattice dynamics simulations as well as scattering rate analysis of phonon transportation. Combining with molecular dynamics simulations, the thermal transport and mechanical properties were systematically investigated. The presented methodology and the inter-atomic potentials pave the way for a systematic approach to model heat transport in SiC related devices using multiscale modelling.

## 1. Introduction

Silicon carbide (SiC) has excellent mechanical, chemical, thermal, and electronic properties, including high stiffness, high thermal conductivity, high melting point, high breakdown electrical field, high carrier saturation speed, low neutron cross-section, low swelling rate, good resistance to wear, oxidation and creep, thermal shock [1,2]. The combination of these advantages makes SiC a promising candidate for a wide-range industrial applications, such as radiation-tolerant structural materials for fusion reactor components [2,3], accident tolerant fuel cladding systems for the advanced nuclear reactors [4], and substrates for the wide bandgap semiconductor to be applicable to high-power switching, high temperature and radiation–resistant electronics [5]. Thermal conductivity is considered as one of important metrics to evaluate the performance of SiC materials under these applications. In particular, the efficient thermal management is usually the key to these high-temperature applications for stable performance and a long lifetime. For example, nuclear grade SiC/SiC composites need to withstand instantaneous high temperature and high-energy neutron-irradiation [3]. Its thermal conductivity changes have an important impact on ensuring the safe and stable operation of nuclear reactors. Therefore, a number of experimental studies [2,6–12], using Raman or TDTR, have been reported to understand thermal transport in SiC materials. However, measurements of thermal conductivity generally suffer from heat losses, nonuniform heating, and errors introduced due to approximations that account for sample size and structure. As an alternative, computational approaches [13–21] not only offer controllable virtual measurements to determine the thermal conductivity, but also offer us the chance to peek the inner-scattering mechanisms and systematically to understand of influencing factors of the thermal transport properties [22].

However, evaluating the lattice thermal conductivity with a predictive model is still a formidable task and there is no universal theory that works at a wide range of temperatures or is able to take into account all possible effects [23]. Lattice vibrations (phonons) contribute mostly to thermal conductivity of the covalent compound SiC, whose electrons are considered as in the ground state adiabatically. The lattice dynamics approach based on the Boltzmann transport equation (BTE) [24–26] and molecular dynamics (MD) simulations have been the two most frequently used computational methods that include atomistic level details in studying thermal transport. The BTE based approach requires the determination of phonon frequencies and group velocities from atomistic level lattice dynamics of an *N*-body system. The MD based approaches include the steady state non-equilibrium molecular dynamics (NEMD) method [27–29], the approach to equilibrium molecular dynamics (AEMD) method [23,30–33], and equilibrium molecular dynamics (EMD) methods (including Green-Kubo formulations [34,35] and Einstein relations [22]), etc. The accuracy of all these methods lies mainly in the exact description of interaction between C, Si atoms, which can be resorted either to first-principles methods such as the density functional theory (DFT) [36,37], or to inter-atomic potentials (IAPs, also called force fields). The first-principles methods are usually accurate but has very small limitation of the system size with fewer atoms (normally less than 1000), due to the high computational complexity. While the traditional empirical IAPs (E-IAPs), where the system energy is an analytical function of the atomic positions, are computationally more efficient to grant access to larger time and length scales but limited by the accuracy and the transferability. Therefore, one accurate IAP is required for the study of the thermal conduction of SiC system.

Currently, it is found that more than 15 different E-IAPs are available for SiC system in the literature, e.g., PT84 [38], T89 [39,40], T90 [41], T94 [42], HG94 [43], HG95 [44], TY95 [45], DS96 [46], DD98 [47], GW02 [48], EA05 [49], VK07 [50], LB10 [51], KE14 [52] and KN17 [53]. Each potential comes with its strengths and weaknesses. PT84 was proposed by Pearson et al [38] by truncation of the Born-Openheimer expansion up to three-body level, and it includes two-body interaction (generalized Lennard-Jones form) and three-body interaction (Axilrod-Teller form [54]). The adjustable parameters are fitted to experiments of bond lengths and sublimation energies of several clusters and solids [38]. HG94 [44] is the modified version of PT84 through giving more accurate fitting for experimental sublimation energy of 3$C$-SiC (β-SiC with cubic structure). PT84 and HG94 are similar to the form of the Stillinger-Weber (SW) [55] potential. In addition, VK07 [50] and LB10 [51] (the form of environment-dependent interaction potential (EDIP) [56]) can also be considered as SW or SW-like potentials. VK07 shows a good performance for 3$C$-SiC with satisfactory reproduction of cohesive energy, elastic constants and melting temperature. LB10 includes two sets of potential parameters (LB10-A and LB10-B), LB10-A is developed with the parametrization of the original EDIP potential of bulk-Si [56], while all the parameters of LB10-B were optimized in the fitting procedure [51]. LB10-B demonstrated a good applicability to point defects, extended defects (dislocations) and high-pressure phases. HG95 and KE14 are the models of the form of the modified embedded atom method (MEAM) [57]. It was found that HG95 can well reproduce the formation and migration energies of point defects and the formation of anti-site defects. KN17 [53] is one of angular-dependent potential (ADP) [58] models and focuses on the description of the deformation and fracture of SiC.

The Tersoff type (or bond order type) [59] potential, which extended the conventional pair potential to describe many-body effects by assuming that the pair potential coefficients depend on the local environment, is widely used to model the covalent interactions in the structure of SiC, e.g., T89 [39,40], T90 [41], T94 [42], TY95 [45], DD98 [47], GW02 [48] and EA05 [49]. T89 begins with the potentials derived earlier for elemental Si [59,60] and C, and parameters describing Si-C interactions are determined from the elemental parameters by an interpolation scheme. The parameters for C of T90 were changed to reproduce the formation energy of C-vacancy with more accuracy than T89, at the expense of a poorer description of graphite. The cut-off distance of T90 is shorter than that of T89. T94 has similar parameters for T90, but use the longer cut-off distance as T89, can reproduce more reasonable heat of mixing. DD98 has been further modified to match *ab initio* calculations for short-range interactions. TY95 is one modified version of T89 by eliminate the unphysical effects associated with large volumetric deformation. EA05 can well reproduce elastic constants and point defect energies. However, there have still been some obvious drawbacks (e.g., the accuracy) for these E-IAPs as described below.

Despite these ongoing efforts of IAP constructions, there have been challenges in computational approaches for the studying of thermal conduction in SiC [15,18], because there has been no accurate IAP constructed with special care about the properties of phonon transportation. As the lattice thermal conductivity of a perfect crystal is mainly due to anharmonic three-phonon processes, directly related to the third derivatives of the potential energy with respect to atomic displacements, and the latter is generally not fitted or considered in the design of the E-IAPs, there is really no good reason to expect an accurate value for the thermal conductivity calculated from the E-IAPs. Essentially, the limited predictive power of the E-IAPs is due to the fixed analytical function form with fewer adjustable empirical parameters and the small size of statical dataset with inappropriate distribution. The dilemma of accuracy-versus-efficiency in the description of interatomic interaction has confronted the molecular simulation community for several decades [61–64].

Thanks to the recent advancement in artificial intelligence and machine learning (ML), new paradigm in atomistic simulations have come forth in recent years [62,65–67], where machine learning methods are adopted to fit IAPs (ML-IAPs) [63,68–71] from datasets calculated by DFT based method. A good ML-IAP can be nearly as computationally efficient as E-IAPs while retaining much of the accuracy of first principles calculations. Several different ML potentials have been proposed to predict thermal properties of crystalline solid such as Si [72,73], Zr [74], graphene [75], $MoS_2$ [76] and $Ga_2O_3$ [77] and so on.

According to the functional forms, ML-IAP models can be classified into two representative styles as the kernel-based models like the Gaussian approximation potential (GAP) [69] and the deep neural network (DNN) based models like the Behler-Parrinello model [68] and the deep potential (DP) model [62–64]. DNN provide an accurate tool for the representation of arbitrary functions, which is advanced in the description of IAPs. The end-to-end based DP model has demonstrated to achieve an accuracy comparable to first-principles calculations and an efficiency close to E-IAPs. An open-source named DeePMD-kit package [61], interfaced with TensorFlow [78] and LAMMPS package [79], has been developed to train the DP model and run DP-based MD simulation. In addition, one integrated software package named DP-GEN [80], in the framework of concurrent learning [81], has been developed to generate adaptive datasets for the training of DP models. Therefore, DP models have already been used to handle a wide variety of problems [82–87] from various disciplines, some of which are intractable by traditional methods.

In this work, we develop two DP models for SiC based on two adaptively generated datasets within the density functional approximations at the local density and the generalized gradient levels, with particular focus on phonon behaviours and thermal properties. These two DP-IAPs can well reproduce the structural properties and thermal properties with high accuracy. The rest of the paper is organized as follows. In Sec. 2, we introduce the construction of the DP-IAPs, including the mathematical structure of DP model, dataset generation workflow, the DFT calculation details and training details. The computational methods employed in this work, including the methods for structural properties and the methods for thermal properties (e.g., harmonic approximation method, quasi-harmonic approximation (QHA) and Boltzmann transport equation (BTE)), are briefly described in Sec. 3. In Sec. 4, we will demonstrate the validity of the obtained DP-IAPs by comparing the results of structural properties with the experiment or with *ab initio* calculations, and then use lattice dynamics, QHA and BTE methods to investigate the thermal properties of SiC with the DP-IAPs. Finally, the summaries and conclusions are presented in Sec. 5.

## 2. Construction of DP-IAPs

Developing DP-IAPs involves two components: model construction and dataset generation. In this work, we use the DeePMD-kit package [62] for training the DP models, LAMMPS package [79] for MD simulations and VASP package [88–90] for *ab initio* calculations.

## 2.1 Mathematical structure of DP model

In DP model, the total potential energy of each structure can be decomposed into energy contributions from constitute atoms, $E=\Sigma_i E_i$ ($i$=1, 2, …, $N$). The atomic energy ($E_i$) depends only on the chemical species ($\alpha_i$) and the local environment ($\mathcal{R}^i$) of atom $i$. The local environment ($\mathcal{R}^i$) is composed of the relative coordinates ($\boldsymbol{r}_{ji}=\boldsymbol{r}_j-\boldsymbol{r}_i$) of the near neighbours ($j$) of atom $i$ within a cut-off radius ($R_c$). The sub-network for $E_i$ consists of an encoding neural network (ENN) and a fitting neural network (FNN). The ENN is specially designed to map the local environment ($\mathcal{R}^i$) to an embedded feature space, which preserves the translational, rotational, and permutational symmetries of the system. The FNN is a fairly standard fully-connected feedforward neural network with skip connections, which maps the embedded features to an "atomic energy". In this work, a three-layer ENN of size (25, 50 and 100 nodes), following a ResNet-like architecture [91], was use to map the original inputs (the chemical species and the atomic coordinates) to symmetry-preserving components (the encoding feature matrix, $\mathcal{D}^i$). In addition, the FNN is a three-layer feedforward neural network containing 240 nodes in each layer. The projection dimension is set to be 12. More details about this scheme and related models can be found in the original papers of DP model [61,63].

## 2.2 Dataset generation

In order to train DP models, the reference datasets, typically consisting of the snapshot of the atomic configuration (atomic coordinates, atomic species, and cell tensors) and the labels (energies, forces, and/or virial tensors), should be firstly constructed by DFT method. In this work, the reference datasets were generated according to the workflow of DP-GEN package [80], which contains a series of successive iterations (exploration, labelling and training). Before the iterations, the initial datasets were prepared by *ab initio* MD (AIMD) simulations, which are conducted at 300K for the distorted supercells of various SiC polytypes. In this work, we focus on the most abundant polytypes of SiC, i.e., 3*C*-SiC (zinc blende, β-SiC), 2*H*-SiC (wurtzite), 4*H*-SiC and 6*H*-SiC, whose structural differences are the stacking sequence of Si-C bilayers along the direction perpendicular to the closed-packed plane. Those supercells, including 32~64 atoms, were compressed uniformly with a scaling factor *α* ranging from 0.96 to 1.06. Then four coarse DP models with different initial random numbers are trained according to the initial datasets, which including the structures of an isolated atom (C and Si).

After that we enter into the master process (i.e., the successive iterations), where the data is accumulated on the fly as it proceeds. During the exploration, a lot of configuration snapshots were generated in MD simulations based on these coarse DP models for various supercells with different initial configurations (50 randomly-perturbed supercells for each SiC polytype). The MD simulations were conducted using isothermal-isobaric (NPT) ensemble under different thermodynamic conditions (50K~1320K and 1~50000 Bar). Those configuration snapshots along MD trajectories are recorded at time intervals of $\Delta t$=5~25fs. Then the model deviation ($\varepsilon$), the maximal standard deviation of the atomic force predicted by these coarse DP models, will then be calculated for each snapshot. only those structures whose model deviations ($\varepsilon$) falls between 0.15 and 0.3 eV/Å will be considered candidates for labelling procedure, in which accurate *ab initio* energies, forces and virial tensors are calculated and added to the training dataset. Over $1\times10^7$ structures were explored in the whole concurrent learning process, and about 0.01% of them were finally labelled. After labelling, four new coarse DP models are obtained by training with the accumulated training dataset. These concurrent learning steps were iteratively repeated until convergence is achieved, i.e., the configuration space has been explored sufficiently, and a representative dataset with a finite number of structures has been accurately labelled. At last, a productive DP model is trained with the sufficient dataset.

**2.3 DFT calculation details**

All the energies, atomic forces and virial tensors of the structures in the reference dataset were labelled by DFT calculations. Two types of exchange-correlation functionals, the generalized gradient approximation (GGA) in Perdew-Burke-Ernzerhof (PBE) form [92] and the local density approximation (LDA)[93], were used in this work, since they give different phonon-dispersion relations. The four $2s^22p^2$ electrons for C and four $3s^23p^2$ electrons for Si were treated as valence electrons with the core electrons accounted for by the projector-augmented wave (PAW) [94] (the PAW potential in VASP5.4 [88–90]). The kinetic energy cutoff for the plane wave [89,90] was set to 1500 eV (LDA) / 1200 eV (PBE) and the Brillouin zone was integrated using Monkhorst-Pack grids [95] with a consistent spacing $h_k$=0.14Å$^{-1}$. A smearing of 0.1 eV by the Gaussian smearing method was applied to help the convergence. The self-consistent-field iteration will stop when the total energy and band structure energy differences between two consecutive steps are smaller than $10^{-7}$ eV.

## 2.4 Training details

The parameters ($w$) contained in the ENN and the FNN in the DP model are trained with the *ab initio* dataset to minimize the loss function, which is a weighted sum of mean square errors in energies, forces and/or virial tensors. The optimization is solved by TensorFlow's implementation of the Adam stochastic gradient descent method [96]. In each training process, the parameters ($w$) were initialized with random numbers as described above. The total number of training steps is 5000000 for the productive DP models, while 400000 for the coarse DP models. The exponentially decaying learning rate is used. For the productive DP models, the learning rate at the $i$th training step is defined as $r_l(i)=0.001\times 0.95^{i/20000}$, where the initial learning rate is 0.001 and the decay rate is 0.95 with a decay step of 20000. And for the coarse DP models, the decay step is 2000. The energy prefactor ($p_e$), force prefactor ($p_f$) and virial prefactor ($p_v$) in the loss function start at 0.02, 1000 and 0.01, and end at 2, 1 and 1, respectively. Since two types of exchange-correlation functionals were used in this work, two DP models (named DPpbe and DPlda) are trained by using these two datasets (PBE and LDA functionals), respectively.

## 3. Computational Methods

Before the validation of these DP-IAPs, some computational methods employed in this work are briefly presented firstly.

**3.1 Structural properties**

The cohesive energy ($E_c$) was defined as the energy difference between the crystalline state ($E_{bulk}$) and the states of isolated atoms (C and Si, $E_{atom}$ (C) and $E_{atom}$ (Si)). In this work, the cohesive energy is typically represented as the energy per atom and the calculation expression is,

$$E_c \left(\text{in } \frac{\text{eV}}{\text{atom}}\right) = \frac{E_{bulk} - N[E_{atom}(\text{C}) + E_{atom}(\text{Si})]/2}{N}, \quad (1)$$

where $N$ is the number of atoms in the supercell of bulk-SiC. For traditional E-IAPs, $E_{atom}$ (C) and $E_{atom}$ (Si) are usually zero. A smaller cohesive energy value indicates that the structure is more stable.

The lattice constants ($a$ and $c$) of various SiC polytypes at equilibrium status can be deduced by optimizing the supercell sizes to gain the lowest cohesive energy. The relationships between cohesive energy and volume ($V$) can be obtained by calculating the $E_c$ of various supercells under different strains by hydrostatic compression/dilation of the equilibrium supercells. In this work, the volume strain with respect to the equilibrium status is considered as -0.15%~0.15%. In effect, the equilibrium lattice constants were also evaluated from the fitted $E_c$-$V$ curves at the points with lowest $E_c$.

Elastic constants measure the proportionality between strain and stress in a crystal, provided that strain is not so large as to violate Hook's law. The 3$C$ structure symmetry of 3$C$-SiC reduces the number of independent elastic constants to three: $c_{11}$, $c_{12}$ and $c_{44}$. While the symmetry of 2$H$-SiC, 4$H$-SiC and 6$H$-SiC reduce the number of independent elastic constants to five: $c_{11}$, $c_{12}$, $c_{13}$, $c_{33}$, $c_{44}$ and $c_{66}$ (=$(c_{11}-c_{12})/2$). In this work, the elastic constant is determined by applying a small strain to an equilibrium supercell with lowest cohesive energy, measuring the energy versus strain, and determining the elastic constants from the curvature of this function at zero strain. After elastic constants of the single crystal are obtained, then the bulk modulus ($B$), Young's modulus, shear modulus and Poisson ratio can be evaluated by applying the Voigt-Reuss-Hill approximations [97–99], e.g., bulk modulus can be obtained as $B=(c_{11}+2c_{12})/3$ for 3$C$-SiC and $B=(2c_{11}+2c_{12}+4c_{13}+c_{33})/9$ for 2$H$-SiC, 4$H$-SiC and 6$H$-SiC.

In addition, the bulk modulus can also be deduced by the $E_c$-$V$ relationship, which can be described by the Murnaghan equation of state (EOS) [100],

$$E_{\rm c}(V) = E_0 + \frac{BV}{B'(B'-1)}\left[B'\left(1-\frac{V_0}{V}\right) + \left(\frac{V_0}{V}\right)^{B'} - 1\right], \qquad (2)$$

where $E_0$ is the cohesive energy of the supercell at the equilibrium status with volume of $V_0$, $B$ is bulk modulus, and $B'$ is the first derivative of the bulk modulus with respect to the pressure ($B'=\partial B/\partial p$). Therefore, these parameters ($E_0$, $V_0$, $B$ and $B'$) can be obtained by fitting $E$-$V$ curves with the Murnaghan's EOS.

**3.2 Phonon calculations**

The thermal properties can be evaluated from lattice dynamics (phonon calculation). The equilibrium supercell was firstly obtained by minimizing residual forces and optimizing stress tensors based on DFT methods or IAPs. Under the conventional harmonic approximation, all terms in cubic or higher order in the Taylor's expansion of the potential energy are assumed negligible, therefore the crystal potential energy ($E$) is presumed to be a quadratic function of the displacements ($u$) with respect to the equilibrium positions of the atoms,

$$E = \frac{1}{2}\sum_{ll'kk'} u^{\mathrm{T}}(lk)\Phi(lk,l'k')u(l'k'), \tag{3}$$

where $l$ ($l'$) and $k$ ($k'$) are the labels of unit cells and atoms in each unit cell, respectively. In this equation, $\Phi$ is the 2$^{\mathrm{nd}}$ interatomic force constants (IFCs), which can be determined from the atomic forces (Hellmann-Feynman forces in DFT calculations) induced by the displacement of an atom in the equilibrium supercell. The dynamical matrix $D$ is expressed as,

$$D(kk',\boldsymbol{q}) = \frac{1}{\sqrt{m_k m_{k'}}}\sum_{l'} \Phi(0k,l'k')\exp\left(i\boldsymbol{q}\cdot(\boldsymbol{r}(l'k') - \boldsymbol{r}(0k))\right). \tag{4}$$

Then the eigenvalues ($\omega$) and eigenvectors ($A$) can be obtained by diagonalization of the dynamical matrix. A complete solution of the eigenproblem in Eq. (4) is sought in terms of $\omega(\boldsymbol{q})$ and $A(\boldsymbol{q})$, which called the phonon dispersion relations. Then the density of states (DOS, $g$) of phonons can be obtained and defined as [101],

$$g(\omega) = \frac{1}{N_u}\sum_{\boldsymbol{q},\nu} \delta(\omega - \omega(\boldsymbol{q},\nu)), \tag{5}$$

where $\nu$ is mode label (band index), $\delta(\ )$ is Dirac distribution function, $N_u$ is the number of unit cells in crystal. Thus, $g(\omega)$ is normalized as the integral over $\omega$ becomes $3n_a$ since the above equation is divided by $N_u$, where $n_a$ is the atomic number in the unit cell.

2. Based on the conventional harmonic approximation, the energy ($E_{ph}$) of phonon system under Bose-Einstein distribution ($n_0(\omega)=(\exp(\hbar\omega/k_bT)-1)^{-1}$, where $n_0$ is the occupation number of the phonon with $\omega$) can be given from the known phonon frequencies ($\omega$) over Brillouin zone,

$$E_{\text{ph}} = \frac{1}{2}\sum_{q,\nu} \hbar\omega(q,\nu) + \sum_{q,\nu} \frac{\hbar\omega(q,\nu)}{\exp\left(\frac{\hbar\omega(q,\nu)}{k_bT}\right)-1}, \tag{6}$$

where $\hbar$ is reduced Planck constant, $k_b$ is Boltzmann constant and $T$ is temperature. According to the thermodynamic relations, many thermal properties, such as entropy ($S$), Helmholtz free energy ($F$) and heat capacity at constant volume ($C_V$), can be deduced as [101]:

$$S = \frac{1}{T}\sum_{q,\nu} \frac{\hbar\omega(q,\nu)}{\exp\left(\frac{\hbar\omega(q,\nu)}{k_bT}\right)-1} - k_b \sum_{q,\nu} \ln\left[1 - \exp\left(-\frac{\hbar\omega(q,\nu)}{k_bT}\right)\right], \tag{7}$$

$$F = \frac{1}{2}\sum_{q,\nu} \hbar\omega(q,\nu) + k_bT \sum_{q,\nu} \ln\left[1 - \exp\left(-\frac{\hbar\omega(q,\nu)}{k_bT}\right)\right], \tag{8}$$

$$C_V = \sum_{q,\nu} C_{q,\nu} = \sum_{q,\nu} k_b \left[\frac{\hbar\omega(q,\nu)}{k_bT}\right]^2 \frac{\exp\left(\frac{\hbar\omega(q,\nu)}{k_bT}\right)}{\left[\exp\left(\frac{\hbar\omega(q,\nu)}{k_bT}\right)-1\right]^2}, \tag{9}$$

where $C_{q,\nu}$ is the mode contribution to the heat capacity.

In this work, the phonon calculation was carried out using PHONOPY package [101,102]. With corrections of long-range interaction of macroscopic electric field induced by polarization of collective ionic motions near the Γ-point where non-analytical term is added to dynamical matrix [103–105] in the calculations of both IAPs and DFTs.

### 3.3 Quasi-harmonic approximation (QHA)

Within the conventional harmonic approximation many important thermal properties, e.g., thermal expansion, specific heat at constant pressure, temperature dependence of the elastic constants, and phonon frequencies or finite phonon lifetimes cannot be described. However, these temperature-dependent properties of an anharmonic system can be calculated based on the so-called quasi-harmonic approximation (QHA). In the QHA, the 2nd IFCs are renormalized by taking into account an explicit dependence upon the volume, i.e., the harmonic approximation is applied at changed volume near the equilibrium status. In the calculation with QHA, we only focus on 3*C*-SiC since the three axes of 3*C*-SiC are isotropic and convenient for hydrostatic change of volume. In this work, 11 volume points (volume error are ±0.05, ±0.04, ±0.03, ±0.02, ±0.01, and 0) were applied to calculate electronic internal energy $U(V)$ and phonon Helmholtz free energy $F(T, V)$ (i.e., Eq. (8)). Then Gibbs free energy $G(T, p)$ at given temperature $T$ and pressure $p$ can be obtained,

$$G(T,p) = \min_{V}[U(V) + F(T,V) + pV]. \qquad (10)$$

The Vinet EOS [106] was used to fit the *G-V* relations in the minimization, then the Gibbs free energies ($G(T, p)$) can be deduced. The corresponding equilibrium volume, isothermal bulk modulus and the derivative of the entropy with respect to the volume as functions of temperature ($V(T)$, $B(T)$ and $dS/dV(T)$) were obtained simultaneously from the EOS and the polynomial fitting for *S* with respect to volume. Phonon frequency decreases typically with the increase of volume, and the slope of each phonon mode is nearly constant in wide volume range. The normalized slope is defined as mode-Gruneisen parameter ($\gamma_{q,v}$),

$$\gamma_{q,v}(V) = -\frac{V}{\omega(q,v,V)}\frac{\partial \omega(q,v,V)}{\partial V}, \qquad (11)$$

which can be used to evaluate the strength of phonon-phonon scattering and then determine phonon life time. The macroscopic Gruneisen parameter ($\gamma$) can be averaged by,

$$\gamma = \frac{1}{C_V}\Sigma_{q,v}\gamma_{q,v}C_{q,v}. \qquad (12)$$

The heat capacity at constant pressure ($C_p$) can be derived from $G(T, p)$ [107],

$$C_p(T,p) = -T\frac{\partial^2 G(T,p)}{\partial T^2} = T\frac{\partial V(T,p)}{\partial T}\frac{\partial S(T,V)}{\partial V}\bigg|_{V=V(T,p)} + C_V(T,V(T,p)), \quad (13)$$

where $V(T,p)$ is the equilibrium volume at $T$ and $p$. The thermal expansion coefficient is an important design parameter for high-temperature application. The coefficient of volumetric thermal expansion ($\alpha_V$) can be defined as,

$$\alpha_V(T) = \frac{1}{V(T)}\frac{\partial V(T)}{\partial T}, \quad (14)$$

which can be deduced from the $V$-$T$ relations. For the cubic structure, the coefficient of linear thermal expansion (CLTE, $\alpha_L$) is approximately equal to,

$$\alpha_L(T) \cong \frac{1}{3}\alpha_V(T). \quad (15)$$

More details about the QHA method can be found in the Refs. [101,107,108].

### 3.4 Boltzmann transport equation (BTE)

In order to obtain the thermal conduction properties, the anharmonic interaction should be considered. The harmonic and anharmonic IFCs can be obtained base on the 2$^{nd}$ and 3$^{rd}$ order derivatives of the potential energy, respectively. The maximum distance among 5-th neighbors corresponding to anharmonic IFCs in the supercell is automatically determined and the cutoff distance is set accordingly. By iteratively solving the linearized phonon Boltzmann transport equation (BTE), the lattice thermal conductivity ($\kappa_L$), phonon scattering rate ($\Gamma$) and Gruneisen parameter ($\gamma$, i.e., Eq. (11)) are calculated as implemented in ShengBTE package [25].

The lattice thermal conductivity tensor can be formulated as [109,110],

$$k_L^{\alpha\beta} = \frac{1}{N_q k_b T^2 \Omega} = \sum_\lambda^\infty n_0(n_0+1)(\hbar\omega_\lambda)^2 v_\lambda^\alpha v_\lambda^\beta \tau_\lambda, \quad (16)$$

where $\Omega$ is the volume of the primitive cell, $\alpha,\beta(=x,y,z)$ are the Cartesian components, $N_q$ denotes the number of $q$-points in a grid used to sample the Brillouin zone. $\omega_\lambda$ and $\tau_\lambda$ is the angular frequency and relaxation time respectively, $\lambda$ is the phonon mode which consists of the wave vector ($\boldsymbol{q}$) and the phonon branch ($v$). $v$ is group velocity and $n_0$ is the Bose-Einstein distribution function. The phonon lifetime ($\tau$) is the inverse of scattering rate ($\Gamma$) which is contributed by three-phonon scattering $\Gamma^{\pm}_{\lambda\lambda'\lambda''}$ and isotopic disorder scattering. The effect of isotopic disorder on the phonon scattering rate is computed automatically in the ShengBTE package [25]. The $\Gamma^{\pm}_{\lambda\lambda'\lambda''}$ can be expressed as [25],

$$\Gamma^{\pm}_{\lambda\lambda'\lambda''} = \frac{\hbar\pi}{4}\begin{Bmatrix}n'-n''\\n'+n''+1\end{Bmatrix}\frac{\delta(\omega\pm\omega'-\omega'')}{\omega\omega'\omega''}|V^{\pm}_{\lambda\lambda'\lambda''}|^2, \qquad (17)$$

where the upper (lower) row in the curly brackets follows the + (−) sign, which corresponds to the absorption (emission) processes. $n'$ stands for $n^0_{\lambda'}$, and so forth. $V^{\pm}_{\lambda\lambda'\lambda''}$ is the scattering matrix elements given by [110,111],

$$V^{\pm}_{\lambda\lambda'\lambda''} = \sum_{i\in \text{u.c.}}\sum_{j,k}\sum_{\alpha\beta\gamma}\Phi^{\alpha\beta\gamma}_{ijk}\frac{e^{\alpha}_{\lambda}(i)e^{\beta}_{v',\pm\mathbf{q}'}(j)e^{\gamma}_{v'',-\mathbf{q}''}(k)}{\sqrt{m_im_jm_k}}, \qquad (18)$$

where $\Phi^{\alpha\beta\gamma}_{ijk}$ is the anharmonic IFCs, $\mathbf{e}_{v,\mathbf{q}}$ is the normalized eigenvalues of the three phonons involved. $i$, $j$ and $k$ run over atomic indices and $i$ only runs over the atoms in the central unit cell, but $j$ and $k$ cover the whole system, $\alpha$, $\beta$ and $\gamma$ indicate the Cartesian components. $m_i$ denotes the mass of the $i$-th atom and $e^{\alpha}_{\lambda}(i)$ is the $\alpha$ component of the eigenvalue of mode $\lambda$ at the $i$-th atom.

For the DFT IFCs, the VASP package [88–90] is used. The pseudopotentials and cut-off energy are the same with the training process of DP models. A 5×5×5 supercell is used in the 3$^{\text{rd}}$ IFCs calculations. To reduce the total displaced supercells, the compressive sensing lattice dynamics (CSLD) [112] method is used to obtain the IFCs. When solving the BTE, we performed convergence tests concerning the $\boldsymbol{q}$ mesh. Tests using $q$ meshes of 40 × 40 × 40 and 35 × 35 × 35 give less than 0.5% differences. For convenience, we choose the latter as $q$ mesh size. For the MD IFCs, the LAMMPS package [79] is used to calculate the interatomic forces, and the IFCs are obtained by our in-house code. Finally, the ShengBTE package [25] is used to solve the Boltzmann transport equation iteratively.

## 4. Results and discussion

### 7.1 4.1 Performance of the DP models

Fig. 1 compares the predicted energy, force and virial values from DP models with the corresponding *ab initio* datasets. The figures show that predictions of the DP models are highly accurate. Based on these data, the mean squares root of the differences (RMS-d=$\sqrt{\frac{1}{N}\sum_{i=1}^{N}(y_{\text{DP},i} - y_{\text{DFT},i})^2}$, where $y$ is the data and $N$ is the number of the data) of energy, force and virial energy obtained by DP models and DFT methods, were calculated as given in panels of Fig. 1. For DPlda model, the errors (RMS-d values) are 1.32 meV/atom, 0.05 eV/Å and 8.47 meV/atom, respectively. For DPpbe model, the errors are 3.73 meV/atom, 0.12 eV/Å and 19.27 meV/atom, respectively. The error levels are comparable to the results reported by Zhang et al. [62]. In order to further validate the performance of the DP models, we test it with structural properties (cohesive energy, lattice constant, bulk modulus, elastic constants, and so on), by comparing the results obtained by experiments, DFT calculations and some aforementioned E-IAPs. The results for 3*C*-SiC, 2*H*-SiC, 4*H*-SiC and 6*H*-SiC are shown in Tables 1 to 4, respectively.

The experimental data for $E_c$ of 3$C$-SiC can be found in the literatures [38,113] as shown in Table 1 (6.34 or 6.43 eV/atom). In our knowledge, there is no reported experimental data for $E_c$ of non-cubic structures (except 3$C$-SiC). First-principles calculations [53,114], especially for LDA based method, tend to overestimate $E_c$ for 3$C$-SiC with 6.66 ~ 7.53 eV/atom range. The PBE based DFT calculation in this work exhibits better agreement with the experimental data for $E_c$ of 3$C$-SiC with variations of less than 0.01 eV/atom [38]. And for DPlda model, $E_c$ is 7.38 eV/atom, in good agreement with LDA calculation in present work. For DPpbe model, $E_c$ is 6.432 eV/atom, in good agreement with experiment [38] and PBE calculation in present work. This level of precision was also reached in previous potential of E-IAPs [38,39,50–53,115,41,42,44–49], because its target value of $E_c$ of fitting is usually using experimental data. And in this case, They can reproduce $E_c$ with error of -0.18~0.1eV by comparing the experimental data, except that PT84 [38] overestimate $E_c$ for 3$C$-SiC.

However, inconsistent energetic orders of these four polytypes occurred in DFT calculations. In LDA based calculations of Refs. [14,116–118] and present work, 4$H$-SiC was found as the most stable structure and then, 6$H$-SiC, 3$C$-SiC and 2$H$-SiC. While in PBE based calculations of Ref [119] and present work, the energetic order is 6$H$-SiC, 4$H$-SiC<3$C$-SiC<2$H$-SiC. This discrepancy stems from the same local tetrahedral environment of the polytypes, as the result the small energy difference between these four phases are at the order of meV/atom, which is close to the training error of the energy and are therefore difficult to reproduce with IAPs. Nonetheless, the DPpbe model can reproduce that 6$H$-SiC has the lowest energy and 2$H$-SiC has the highest energy. As shown in Table 1, the experimental lattice constant ($a$) of 3$C$-SiC is 4.36 Å [14,38,120,121], well reproduced by DFT calculations with 4.308~4.38 Å in Refs. [14,53,114,117,121–124] and our DP-IAPs. The lattice constant is also reproduced by most E-IAPs [39,41,52,53,115,42,45–51] with 4.28~4.411 Å, while PT84 [38] and HG95 [44] is slightly underestimated. In this work, $a$ is 4.331 Å (DPlda) and 4.379 Å (DPpbe), which are good agreement with LDA calculation (4.332 Å) and PBE calculation (4.379 Å), respectively.

Available experimental bulk modulus ($B$) is vast and scattered in the range of 211~322 GPa [2,14,44,120,121,125–130], while the DFT results from Refs. [14,16,53,114,117,122] and this work are concentrated in 210 ~229 GPa. Most E-IAPs can reproduce the reasonable values (211~241 GPa [39,41,42,44–53,115,125]) for $B$, only PT84 [38] and HG94 [43] are overestimated. In this work, $B$ is 229 GPa (DPlda) and 210 GPa (DPpbe), which are good agreement with LDA calculation (229 GPa) and PBE calculation (213 GPa), respectively. Similarly, these two DP-IAPs can also well reproduce these elastic constants ($c_{11}$, $c_{12}$ and $c_{44}$) of 3$C$-SiC as shown in Table 1. It can be seen that 3$C$-SiC likes several other covalently bonded materials (Si, GaAs) exhibits a negative Cauchy discrepancy ($c_{12}$-$c_{44}$<0).

As mentioned above, the main structural difference of these four SiC polytypes are the stacking sequence of the close-packed Si-C bilayer. The ideal 2$H$-SiC follows an ***abab***… stacking, the ideal 3$C$-SiC follows an ***abcabc***… stacking, the ideal 4$H$-SiC follows an ***abacabac***… stacking, and the ideal 6$H$-SiC follows an ***abcacbabcacb***… stacking, where ***a*** (***b*** or ***c***) represents one Si-C bilayer. Thus, the lattice constants ($a$ and $c$) of 2$H$-SiC, 4$H$-SiC and 6$H$-SiC approximately follow the corresponding geometrical relationship ($a_{2H/4H/6H} \approx a_{3C}/\sqrt{2}$, $c/a \approx 1.633$ (2$H$), 3.266 (4$H$) and 4.899 (6$H$)) as shown in Table 2 to 4. It can be seen that lattice constants of 2$H$-SiC, 4H-SiC and 6H-SiC obtained by these two DP-IAPs are good in agreement with experimental values [2,14,116,117,131], DFT results in Refs. [14,53,116,117,121] and present work, and also E-IAP results [39,41,53,42,44,47–52]. In addition, the bulk modulus ($B$) of 2$H$-SiC, 4$H$-SiC and 6$H$-SiC is approximately equal to that of 3$C$-SiC, since these polytypes share the same local tetrahedral environment. The elastic constants ($c_{11}$, $c_{12}$, $c_{13}$, $c_{33}$, $c_{44}$ and $c_{66}$) of 2$H$-SiC, 4$H$-SiC and 6$H$-SiC are also comparable as shown in Table 2 to 4. In brief, our two DP-IAPs can well reproduce the primary structural properties of these SiC polytypes.

The cohesive energy ($E_c$) per atom is plotted as functions of volume ($V$) of 3$C$-SiC for both DP-IAPs and DFT data, as comparison, in Fig. 2 (a). It can be seen that the E-IAPs [39,41,42,44,47–52] usually overestimate the $E_c$ comparing to the DFT results. Moreover, T89 [39,40], T94 [42] and DD98 [47] have significant deviation from DFT calculations especially at lower volume. While the $E_c$-$V$ relationships of these two DP-IAPs are well agreement with that of corresponding DFT calculations. The similar features of $E_c$-$V$ relationship were also found for 2$H$-SiC, 4$H$-SiC and 6$H$-SiC. It can be seen from Fig. 2 (b), taking DPlda and DPpbe for 3$C$-SiC as examples, that Murnaghan's EOS (Eq. (2)) can well describe the $E_c$-$V$ relationship. The fitted $B$ and $B'$, including the results from some E-IAPs [39,41,42,44,47–52], are also shown in parentheses in Table 1 to 4, comparing to the existed data. The fitted bulk modulus ($B$) is very close to the results from the method of small strain. Only a few experimental $B'$ for 3$C$-SiC are available in literatures [50,121,127,129,130] in the range of 2.9~4.1. These two DP-IAPs can well reproduce $B'$, in agreement with the experiment data and the DFT results from Refs. [14,116,121,122] and present work. However, the existing E-IAPs [39,41,42,44,47–52] usually overestimate $B'$, especially for VK07 [50], while LB10-B [51] highly underestimate $B'$.

## 4.2 Evaluation of thermal properties
### 4.2.1 Phonon behaviours

The main objective of these potentials is the precise description of thermal properties of SiC. Phonon dispersion curves demonstrate frequency ($\omega$) versus wave vector ($q$) along high symmetric direction in crystal and carry the information of the 2$^{nd}$ force constants. The density of states (DOS) of phonons in the entire Brillouin zone is an indispensable factor for further investigation of interesting thermodynamic properties such as heat capacity, thermal expansion and thermal conduction. The phonon-dispersion relations were then tested with these two DP-IAPs, by comparing experimental data [124,132,133], the results of DFT calculations and E-IAPs [39,41,42,44,47–52]. The study of phonon-dispersion relation of SiC with IAPs can be tracked back to the work of Vetelino et al [134]. Non-analytical correction is added to the dynamical matrix to correctly reproduce the LO-TO splitting in both DFT and IAPs cases. As shown in Table 5, the dielectric parameters, including dielectric tensor and Born effective charges, were firstly computed using the perturbative approach implemented in VASP package [88–90]. For example, the static dielectric constants ($\varepsilon_\infty$) of 3$C$-SiC, $\varepsilon_{\infty(\text{LDA})}$=6.93 and $\varepsilon_{\infty(\text{PBE})}$=6.99, compare well with the values of experiments and the other DFT works [14,16,135]. The Born effective charges ($Z_B$) of 3$C$-SiC are $Z_{B(\text{Si, LDA})}$=2.71, $Z_{B(\text{C, LDA})}$=-2.71, $Z_{B(\text{Si, PBE})}$=2.69 and $Z_{B(\text{C, LDA})}$=-2.69 and are in agreement with the values of experiments and the other DFT works [16,127,135]. The different SiC polytypes present similar traces of $\varepsilon_\infty$ and $Z_B$ matrixes, though the hexagonal structures have weak anisotropy of $\varepsilon_\infty$ and $Z_B$ as shown in Table 5.

The phonon spectrums of various SiC polytypes at their equilibrium volume are shown in Fig. 3. The high symmetric paths in Brillouin zone are K (3/8 3/4 3/8) – Γ (0 0 0) – X (0 1/2 1/2) – W (1/4 3/4 1/2) – L (1/2 1/2 1/2) – Γ – L – U (1/4 5/8 5/8) – X directions for the FCC structure of 3$C$-SiC, and K (2/3 1/3 0) – Γ (0 0 0) – M (1/2 0 0) – K – H (2/3 1/3 1/2) – A (0 0 1/2) – L (1/2 0 1/2) – M – Γ – A for the HCP structure of 2$H$-SiC, 4$H$-SiC and 6$H$-SiC. The peculiar feature of the dispersion curves of 3$C$-SiC as shown in Fig. 3 (a) is that the LO and TO modes are split at Γ-point because of the polar character of SiC and the mass difference, which is not shown in diamond [108]. The two DP-IAPs, especially DPlda, can well reproduce the phonon spectrums of the corresponding DFT calculations as well as the experimental data [124,132,133,136], while existing E-IAPs [39,41,42,44,47–52] fail to capture that as shown in Fig. 3. Even good agreement is also observed for the X-point, which implies the strength of chemical bonds introduced by significant amounts of charge transfer can also be well considered and reach satisfying accuracy [14,137]. Moreover, a set of "strong modes" in which the Si and C sublattices vibrate against each other. The anisotropy of one of these modes varies in the same way as the *c*/*a* axial ratios [132]. It should be noted that introducing modulated structures [138] into training sets, with specific displacements along normal modes at particular *q*-points, can further improve the accuracy of the description of phonon features, where discrepancy between DP and DFT are relatively big.

The density of states (DOS) of phonon are shown in Fig. 4. LDA based DFT calculation can deduce a gap of 3.5 THz between the highest acoustic phonon frequency and the lowest optical phonon frequency for 3*C*-SiC, in good agreement with the experimental data (3.6 THz) [124]. Such a gap inhibits three-phonon scattering processes involving optical and acoustic modes because of the energy and quasi-momentum conservation requirements, thus leading to longer phonon lifetimes [139]. DPlda can well reproduce the gap of 3.8 THz, while DPpbe slightly underestimate the gap of 3.2 THz though DPpbe can well reproduce the results of PBE calculations (3.3 THz). However, a large deviation can be usually observed in the gap for the existing E-IAPs [39,41,42,44,47–52]. In addition, there exists one small gap (~ 0.9 THz) and pronounced maxima in optical range of the DOS due to the weak dispersion of the LO and TO phonon branches, while these features cannot be captured by the existing E-IAPs [39,41,42,44,47–52]. In shortly, these two DP-IAPs can well reproduce the DOS of the corresponding DFT calculations, while existing E-IAPs [39,41,42,44,47–52] fail to capture the DOS, even though T89 [39,40], GW02 [48], EA05 [49] and VK07 [50] can give the DOS features closer to those of DFT calculations.

### 4.2.2 Thermal dynamics properties

As shown in Fig. 5, the entropy ($S$), Helmholtz free energy ($F$) and heat capacity at constant volume ($C_V$) as functions of the temperature ($T$) are calculated according the Eqs. (7-9) for various SiC polytypes. It can be seen that these two DP-IAPs can well produce the $S$-$T$, $F$-$T$ and $C_V$-$T$ curves of the corresponding DFT calculations. However, the existing E-IAPs [39,41,42,44,47–52] fail to capture that, e.g., the deviation increased as increasing of temperature for $S$ and $F$, while the pronounced deviation was occurred in the low temperature range for $C_V$. The values of $C_V$ for various IAPs and DFTs are close and approach the constant value $3n_a k_b$ (Dulong-petit law) at high temperature. In addition, the entropy data of 3$C$-SiC from the DP-IAPs are also agree well with the experimental data [140] as shown in Fig. 5 (a).

As shown in Fig. 6, these two DP-IAPs can well reproduce the $G$-$T$, $V$-$T$, $B$-$T$, d$S$/d$V$-$T$ and $\gamma$-$T$ relations of the corresponding DFT calculations. The difference of $G$ between LDA (or DPlda) and PBE (or DPpbe) is mainly from the difference of cohesive energy ($E_c$ or $U$) as shown in Table 1. The E-IAPs usually fail to capture these relations. GW02 [48] has faster decline rate of $G$ (see Fig. 6 (a)) than that of DP-IAPs or DFT calculations, and presents thermal contraction feature (see Fig. 6 (b)), which is significantly different that of DP-IAPs or DFT calculations. The $B$-$T$ relations (see Fig. 6 (c)) of T89 [39,40], T94 [42], HG95 [43], DD98 [47], GW02, VK07 [50] and LB10 [51] obviously deviate from the results of DP-IAPs or DFT calculations. The d$S$/d$V$-$T$ relations (see Fig. 6 (d)) of T89, T94, DD98, GW02 and LB10 obviously deviate from the results of DP-IAPs or DFT calculations. The $\gamma$-$T$ relations (see Fig. (e)) of T89, T94, DD98 obviously deviate from the results of DFT calculations and the $\gamma$ of LB10 presents zigzag variation with the increase of temperature. The calculated macroscopic Gruneisen parameter $\gamma$ in the early LDA-QHA work of Malakkal et al [16] is 0.92~0.93 (at 1200~2500K), which is in good agreement with the results of DP-IAPs (0.86~0.88 at 1200~1800K) and DFT calcuations (0.95~0.96 at 1200~1800K) in present work. For the results of DP-IAPs or DFT calculations, the $G$, $V$ and $B$ do not change significantly below room temperature and the noticeable variation is occurred at high temperature range, while the noticeable variation of d$S$/d$V$ and $\gamma$ is occurred below room temperature.

The relative bulk modulus ($R_B=B(T)/B(0)$) at elevated temperatures are shown in Fig. 7, along with the experimental data [2,141–148] originally compiled in Ref. [2]. The $R_B$ of IAPs and DFTs can be easily deduced from the data present in Fig. 6 (c). It can also be seen that comparing to the DFT results, T89 [39,40], T94 [42], DD98 [47] and LB10 [51] usually overestimate the $R_B$, while HG95 [43] and VK07 [50] usually underestimate the $R_B$, especially at high temperature range. However, these two DP-IAPs can well reproduce the $R_B$-$T$ relations of DFT calculations, which are also close to the experimental results [2,141–148], although the experimental data at high temperature are scattered which may due to the different sample state (e.g., the residual silicon and grain boundary in SiC [2]), experimental method and conditions.

According to Eq. (13), the heat capacities at constant pressure ($C_p$) for 3$C$-SiC were calculated at the temperature range 0~1800 K as shown in Fig. 8. It can be seen that these two DP-IAPs can well reproduce the $C_p$-$T$ relations of DFT calculations, which are also in agreement with experimental data [2,6–9,140,149,150], though there exists a slightly underestimation at high temperature range possibly because of the absent contribution of higher-order anharmonic contributions. It is impressive that most E-IAPs [39,41,42,47,49–52] can also reproduce the reasonable $C_p$-$T$ relations, aside from the overestimation of GW02 [48] at low temperature range and underestimation of HG95 [43] at middle temperature range. It should be noted that the difference between $C_p$ and $C_v$ (see Fig. 5 (a)) is usually small especially at low temperature range. The temperature-dependence of the heat capacity can be divided into two temperature zone, a rapid increase at low temperature (~$T^3$, Debye model) and a gradual increase at higher temperature (approach $3n_a k_b$, Dulong-petit law).

According to the QHA method (Eqs. (14-15)), the calculated CLTE ($\alpha_L$) for 3$C$-SiC at the temperature range 0~1800 K were shown in Fig. 9, in comparison with the experimental data [2,6,9–12,151] and other DFT results [108]. Comparison of the results shows that these two DP-IAPs can well reproduce the CLTE of the corresponding DFT calculations, which also are in good agreement with the data of experiments and other DFT calculations. The slight deviation between DP-IAPs /DFT calculations and the experimental values at high temperatures could be due to the higher-order anharmonic contributions, though the experimental values of CLTE are also scattered. However, the existing E-IAPs [39,41,42,44,47–52] usually obviously fail to capture $\alpha_L$-$T$ relations. The CLTE of GW02 [48] is negative and decreased at first, then increased and decreased at last with the increase of temperature. The LB10 [51] underpredict the CLTE, which presents zigzag variation with the increase of temperature. The CLTE of T89 [39,40], T94 [42] and DD98 [47] increased at first and then decreased with the increase of temperature. HG95 [43] and VK07 [50] overestimate the CLTE, while KE14 [52] underestimate it, despite the variation trend is reasonable. Szpunar et al [152] have also found that the temperature dependence on thermal expansion described by T89 and GW02 does not agree well with DFT results. Only T90 [41] and EA05 [49] in the considered E-IAPs are able to reproduce the CLTE. It should be noted that the temperature-dependence of the CLTE can also divided into two temperature zone, a rapid increase at low temperature and a gradual increase at higher temperature. The CLTE of SiC is significantly dependent on the polytypes, while there is no distinguished difference of the heat capacities for different SiC polytypes.

4.2.3 **Thermal conductivity**

In the following, we demonstrated the accuracy of the fitted DP models, which can capture the thermal conduction properties of SiC from the aspects of phonon behaviour. We calculated the thermal property of 3$C$-SiC with the BTE based on the IFCs from DP-IAPs and the DFT calculations, respectively. The lattice thermal conductivity ($\kappa_L$), phonon scattering rate ($\Gamma$) and Gruneisen parameter ($\gamma$) are obtained. The calculated $\kappa_L$ of bulk 3$C$-SiC, including the isotopic effect is shown in Fig. 10 along with experimental results [2,9,153–156]. At 300 K the values of $\kappa_L$ are 405.06 W/m K, 502.66 W/m K, 445.16 W/m K, 493.63 W/m K, 1031.9 W/m K and 1035.7 W/m K for the calculations of DPpbe, DPlda, PBE, LDA, EA05 [49] and T90 [41], respectively. The results of DP-IAPs match with the results of the DFT very well over the entire temperature range, and are in good agreement with the experimental values [2,9,153–156] at higher temperatures, indicating that the DP-IAPs based on either LDA or GGA pseudopotential can accurately calculate the $\kappa_L$ of SiC. In contrast, the results of traditional E-IAPs greatly differ from the DFT calculation results and experimental values, especially at higher temperatures, which highlights the accuracy of our DP-IAPs. Our theoretical calculation results to be higher than the experimental value at low temperature, which may because the extra scattering of phonon from defects such as grain boundaries in the experiment sample. Thus the Slack's experiments [154] of high-purity SiC have the highest thermal conductivity in all experiments [2,9,153–156]. At elevated temperatures, where phonon-phonon scattering dominates heat transport in intrinsic semiconductors, the calculated results match well with the experimental values [2,9,153–156].

Phonon lifetime $\tau$ is another important parameter for evaluation of thermal conductivity, which are shown in Fig. 11 for DPpbe, DPlda, PBE, LDA, EA05 [49] and T90 [41]. The corresponding Gruneisen parameter $\gamma$ is shown in Fig. 12, for different branches of phonon as function of frequency of phonon at different $q$-points. We find that the acoustic phonons, especially low-frequency phonons, have the lowest scattering rates and therefore contribute the most to the total thermal conductivity. The results of phonon lifetime of the DP-IAPs and the DFT calculation are in good agreement, with a deviation in the range of 10 THz~14 THz. The $\tau$ of DFT calculation has a small jump but the DP-IAPs results do not. The results of Gruneisen parameter the DP-IAPs and the DFT calculation are roughly in good agreement. In sharp contrast, the calculation results of traditional E-IAPs significantly deviate from the DFT results.

In summary, the fitted DP potentials can accurately calculate the thermal properties of SiC, which has a huge advantage over the traditional E-IAPs.

## 3. Summary and conclusion

We have proposed two deep potential models (DPlda and DPpbe) for molecular dynamics simulations of SiC based on two adaptively generated datasets (LDA and PBE based DFT calculations) using the deep neural network methods. The present DP-IAPs accurately reproduce the structural properties, phonon behaviour and thermal properties of various SiC polytypes, while the existing conventional IAPs usually fail to describe that, especially for phonon behaviour and thermal properties. Comparison of these two DP-IAPs, DPpbe can give more accurate structural properties, while DPlda can reproduce the phonon spectrums, DOS features and other thermal properties closer to the experimental results, which is implying that our DP-IAPs can not only describe the second order force constants which strongly rely the precise description of interatomic bonding, but also can precisely capture the third order force constants which essentially reflect the surrounding bonding characteristics for each atoms with information from both bond lengths but also with different angles. Therefore, we are confident on the application of our SiC DP-IAPs in future on-chip simulation not only in thermal conduction but also in the situation where the desired properties are sensitive to those atomic environments.

**Acknowledgements**

This study was sponsored by National Natural Science Foundation of China

**Table 1.** Structural properties of 3$C$-SiC: cohesive energy $E_c$ (in eV/atom), lattice constant $a$ (in Å), bulk modulus $B$ (in GPa), elastic constants $c_{11}$, $c_{12}$, and $c_{14}$ (in GPa), and the derivative of the bulk modulus with respect to the pressure $B'$. The data derived from Murnaghan equation of state [100] are given in parentheses and the data calculated in this work are displayed in bold.

| 3$C$-SiC | DPlda | DPpbe | E-IAPs | DFT | Experiment |
|---|---|---|---|---|---|
| $E_c$ | 7.380 | 6.432 | 7.705[a], **6.165**[b], 6.18[c,d], 6.4[e], 6.21[f], 6.434[g], 6.34[h,i,j], 6.412[k], 6.359[l], 6.338[m], 7.55[n], **6.434**[o], **6.375**[p] | **6.432**[t], **7.379**[u], 6.66[v], 7.53[w] | 6.34[af], 6.43[ag] |
| $a$ | 4.331 | 4.379 | 4.18[a], **4.321**[b], 4.32[c], 4.36[d,k], 4.2[e], 4.307[f], 4.28[g], 4.359[h], 4.349[i], 4.358[j], 4.411[l], 4.364[m], 4.381[n], **4.280**[o], **4.359**[p] | **4.379**[t], **4.332**[u], 4.361[v], 4.376[w], 4.344[x], 4.36[y], 4.38[z], 4.308[aa], 4.328[ab], 4.343[ac] | 4.36[ag, ah, ai, aj] |
| $B$ | 229 (227) | 210 (210) | 990[a], **224**[b], 225 (219)[c], 229[d], 211 (**209**)[e], 231 (**229**)[f], 241[g], 224 (**222**)[h], 224[i], 225.2 (**227**)[j], 235[k], 224[l], 226 (**227**)[m], 230[n], **241**[o], **230** (**228**)[p], 225[q], 1183[r], (**223**)[s] | **213** (**211**)[t], **229** (**227**)[u], 212[v], 214[w], 219[x], 210[y], 215[ab], 223[ad] | 220[ah], 224[ai, ak, al], 227[aj], 211[am], 321.9[an], 225[ao], 260[ap], 230[aq], 255.7[ar] |
| $c_{11}$ | 403 | 389 | 1095[a], **437**[b], 436[c], 372[d], **402**[e], 426[f], 447[g], 382[h], 243[i], 390[j], 254[k], 437[l], 394[m], 508[n], **446**[o], **397**[p], 371[q] | **384**[t], **403**[u], 385[w], 390[x], 369[ab], 397[ad] | 360[ah], 390[ai, ao], 363[al], 511[ar], 352.3[as] |
| $c_{12}$ | 142 | 120 | 937[a], **118**[b], 120[c], 157[d], **116**[e], 134[f,g], 145[h], 215[i], 142.6[j], 225[k], 117[l], 142[m], 91[n], **138**[o], **147**[p], 169[q] | **127**[t], **142**[u], 128[w], 134[x], 118[ab], 136[ad] | 150[ah], 142[ai, ao], 154[al], 128[ar], 140.4[as] |
| $c_{44}$ | 254 | 240 | 606[a], **257**[b], 255[c], 256[d], 205[e], **215**[e], 280[f], 293[g], 240[h], 62[i], 191[j], 66[k], 195[l], 168[m], 269[n], **220**[o], **136**[p], 176[q] | **241**[t], **255**[u], 264[w], 253[x], 226[ab], 275[ad] | 150[ah], 256[ai, ao], 149[al], 191[ar], 232.9[as] |
| $B'$ | (3.79) | (3.82) | 4.11[c], (**3.82**)[e], (**4.01**)[f], 4.16 (**4.07**)[h], 5.5 (**6.20**)[j], (0.61)[m], (**3.96**)[p], (**4.06**)[s] | (**3.83**)[t], (**3.78**)[u], 3.88[x], 3.71[y], 3.3[ae] | 3.57[aj], 3.43[an], 2.9[ap], 4.0[aq], 4.1[at] |

**Table 2.** Structural properties of 2*H*-SiC: the energy relative to 3*C*-SiC ($E_c$ (2*H*-3*C*) in meV/atom), lattice constants $a$ and $c$ (in Å), bulk modulus $B$ (in GPa), elastic constants $c_{11}$, $c_{12}$, $c_{13}$, $c_{33}$, $c_{44}$, and $c_{66}$ (=($c_{11}$-$c_{12}$)/2) (in GPa), and the derivative of the bulk modulus with respect to the pressure $B'$. The data derived from Murnaghan equation of state [100] are given in parentheses and the data calculated in this work are displayed in bold.

| 2*H*-SiC | DPlda | DPpbe | E-IAPs | DFT | experiment |
|---|---|---|---|---|---|
| $E_c$ (2*H*-3*C*) | 2.28 | 2.20 | **8.5**[a], 7.0[b], 5.9[c], 6.1[d], 6.7[e], 22[f], **0.0**[g, h, i, j] | **2.91**[k], **2.34**[l], 2.7[m], 2.45[o], 1.8[p], 1.1[q] | |
| $a$ | 3.058 | 3.090 | **3.056**[a], **3.055**[b], **3.026**[c], **3.083**[d, e], 3.061[f], **3.103**[g], **3.085**[h], **3.046**[i], **3.026**[j] | **3.091**[k], **3.058**[l], 3.065[n], 3.069[o], 3.072[p] | 3.07[r], 3.081[s], 3.079[t] |
| $c$ | 5.017 | 5.073 | **5.119**[a], **4.990**[b], **4.942**[c], **5.034**[d, e], 5.023[f], **5.067**[g], **5.037**[h], **4.973**[i], **4.942**[j] | **5.072**[k], **5.020**[l], 5.039[n], 5.034[o], 5.041[p] | 5.042[r], 5.031[s], 5.053[t] |
| $B$ | 230 (228) | 214 (212) | **229 (227)**[a], **224**[b], **241**[c], **211 (209)**[d], **224 (222)**[e], 221.5 **236 (225)**[f], **224 (223)**[g], **226 (227)**[h], **231 (229)**[i], **241**[j] | **214 (212)**[k], **230 (228)**[l], 225[n], 224[o], 215[p] | 223[u] |
| $c_{11}$ | 522 | 485 | **472**[a], **523**[b], **506**[c], **467**[d], **487**[e], 415[f], **293**[g], **437**[h], **483**[i], **506**[j] | **498**[k], **522**[l] | |
| $c_{12}$ | 118 | 106 | **220**[a], **96**[b], **123**[c], **98**[d], **121**[e], 158[f], 207[g], **129**[h], **119**[i], **123**[j] | **96**[k], **114**[l] | |
| $c_{13}$ | 56 | 48 | **1.8**[a], **53**[b], **94**[c], **68**[d], **64**[e], 151[f], 174[g], **112**[h], **92**[i], **94**[j] | **51**[k], **60**[l] | |
| $c_{33}$ | 566 | 547 | **674**[a], **566**[b], **534**[c], **498**[d], **544**[e], 376[f], 326[g], **453**[h], **510**[i], **534**[j] | **525**[k], **557**[l] | |
| $c_{44}$ | 160 | 162 | **104**[a], **192**[b], **176**[c], **167**[d], **160**[e], 127[f], 38[g], **142**[h], **167**[i], **176**[j] | **152**[k], **155**[l] | |
| $c_{66}$ | 202 | 189 | **126**[a], **214**[b], **191**[c], **185**[d], **183**[e], 129[f], 43[g], **154**[h], **181**[i], **191**[j] | **200**[k], **206**[l] | |
| $B'$ | (3.77) | (3.89) | **(3.95)**[a], **(3.82)**[d], **(4.07)**[e], 6.9 **(6.63)**[f], **(4.06)**[g], **(0.61)**[h], **(4.01)**[i] | **(3.86)**[k], **(3.80)**[l], 3.78[n], 3.75[o], 4.2[p] | |

**Table 3**. Structural properties of $4H$-SiC: the energy relative to $3C$-SiC ($E_c$ ($4H$-$3C$) in meV/atom), lattice constants $a$ and $c$ (in Å), bulk modulus $B$ (in GPa), elastic constants $c_{11}$, $c_{12}$, $c_{13}$, $c_{33}$, $c_{44}$, and $c_{66}$ ($=(c_{11}-c_{12})/2$) (in GPa), and the derivative of the bulk modulus with respect to the pressure $B'$. The data derived from Murnaghan equation of state[100] are given in parentheses and the data calculated in this work are displayed in bold.

| $4H$-SiC | DPlda | DPpbe | E-IAPs | DFT | experiment |
|---|---|---|---|---|---|
| $E_c$ ($4H$-$3C$) | 0.61 | 0.13 | 5.8[a], 4.9[b], 5.0[c], 5.6[d], 0.0[e,g], **5.9**[f], 30[h], **0.0**[i,j,k], **14.6**[l] | -0.65[m], **-1.25**[n], -1.2[o], -1.9[p], -1.36[q], -2.5[r], -1.8[s], 0.0[t] | |
| $a$ | 3.06 | 3.094 | **3.055**[a], **3.026**[b], **3.083**[c,d], 3.119[e], **3.062**[f], 3.085[g], 3.092[h], **3.103**[i], **3.046**[j], **3.026**[k], **3.074**[l] | 3.093[m], **3.06**[n], 3.075[q], 3.069[r], 3.067[s], 3.091[t], 3.968[u] | 3.07[v], 3.081[w], 3.073[x] |
| $c$ | 10.018 | 10.133 | **9.979**[a], **9.883**[b], **10.068**[c], 10.067[d], 10.186[e], **10.199**[f], 10.074[g], 10.101[h], **10.135**[i], **9.947**[j], **9.883**[k], **10.063**[l] | **10.127**[m], **10.017**[n], 10.054[q], 10.103[r], 10.068[s], 10.124[t], 10.051[u] | 10.04[v], 10.061[w], 10.052[x] |
| $B$ | 229 (228) | 213 (212) | **224**[a], **241**[b], 211 (209)[c], 224 (222)[d], 223[e], 230 (228)[f], 226 (227)[g], 211[h], **224 (223)**[i], **231 (229)**[j], **241**[k], **238 (219)**[l] | 213 (211)[m], 230 (228)[n], 223[q], 218[r], 204[s], 214[t] | 220[y] |
| $c_{11}$ | 515 | 483 | **523**[a], **507**[b], 467[c], 485[d], 536[e], **449**[f], 437[g], 507[h], 290[i], **483**[j], **507**[k], **410**[l] | 498[m], **522**[n], 379[s], 536[t] | 501[y], 507[z] |
| $c_{12}$ | 121 | 104 | 96[a], **122**[b], 98[c], 123[d], 87[e], 193[f], 129[g], 101[h], 210[i], **118**[j], **122**[k], **145**[l] | 94[m], 110[n], 116.5[s], 80[t] | 111[y], 108[z] |
| $c_{13}$ | 57 | 50 | 53[a], **94**[b], 68[c], 64[d], 54[e], **51**[f], 112[g], 43[h], 174[i], **92**[j], **94**[k], **149**[l] | **50**[m], **59**[n], 31[t] | 52[y] |
| $c_{33}$ | 563 | 541 | **566**[a], **534**[b], 498[c], 544[d], 568[e], **576**[f], 453[g], 519[h], 326[i], **510**[j], **534**[k], **439**[l] | 530[m], 568[n], 570[t] | 553[y], 547[z] |
| $c_{44}$ | 166 | 166 | **192**[a], **176**[b], 167[c,j], 160[d], 194[e], **107**[f], 142[g], 224[h], **38**[i], **176**[k], **128**[l] | 158[m], **162**[n], 242[s], 164[t] | 163[y], 159[z] |
| $c_{66}$ | 197 | 189 | **213**[a], **192**[b], 184[c], 181[d], 224.5[e], **128**[f], 154[g], 203[h], **40**[i], **182**[j], **192**[k], **131**[l] | **193**[m], **200**[n], 228[t] | 195[y], 200[z] |
| $B'$ | (3.75) | (3.75) | (3.82)[c], (4.07)[d], (3.95)[f], (0.61)[g], (4.06)[i], (4.01)[j], (6.25)[l] | (3.83)[m], (3.78)[n], 3.66[q], 3.8[r], 3.72[u] | |

**Table 4**. Structural properties of 6$H$-SiC: the energy relative to 3$C$-SiC ($E_c$ (6$H$-3$C$) in meV/atom), lattice constants $a$ and $c$ (in Å), bulk modulus $B$ (in GPa), elastic constants $c_{11}$, $c_{12}$, $c_{13}$, $c_{33}$, $c_{44}$, and $c_{66}$ (=$(c_{11}-c_{12})/2$) (in GPa), and the derivative of the bulk modulus with respect to the pressure $B'$. The data derived from Murnaghan equation of state [100] are given in parentheses and the data calculated in this work are displayed in bold.

| 6$H$-SiC | DPlda | DPpbe | E-IAPs | DFT | experiment |
|---|---|---|---|---|---|
| $E_c$ (6$H$-3$C$) | 0.329 | -0.04 | 4.3[a], 0.2[b, c, d, e], 0.0[f, g], **0.0**[h, i, j], **9.6**[k] | **-0.66**[l], **-1.1**[m], -1.8[n], -1.6[o], -1.5[p], -1.05[q] | |
| $a$ | 3.061 | 3.095 | 3.067[a], 3.055[b], 3.026[c, j], **3.083**[d, e], 3.119[f], 3.085[g], **3.103**[h], 3.046[i], **3.077**[k] | 3.093[l], **3.061**[m], 3.077[n], 3.074[o] | 3.081[r, s], 3.073[t], |
| $c$ | 15.018 | 15.187 | **15.246**[a], **14.969**[b], **14.825**[c, j], 15.102[d], 15.101[e], 15.279[f], 15.112[g], **15.202**[h], 14.920[i], **15.096**[k] | **15.180**[l], **15.018**[m], 15.108[n], 15.1[o] | 15.092[r], 15.12[s], 15.079[t], |
| $B$ | 229 (228) | 212 (212) | 230 (228)[a], 224[b], 241[c, j], 211 (209)[d], 224 (222)[e], 226 (227)[g], 224 (223)[h], 231 (229)[i], 234 (222)[k] | 213 (211)[l], 229 (228)[m], 204[n, o] | 221[u], 211[v, x] |
| $c_{11}$ | 512 | 482 | 431[a], 523[b], 507[c, j], 467[d], 485[e], 436[g], 289[h], 483[i], 407[k] | 492[l], 524[m], 376[o] | 501[u], 464.5[v], 502[w], 479[x], |
| $c_{12}$ | 120 | 102 | 173[a], 96[b], 122[c, j], 98[d], 123[e], 129[g], 211[h], 118[i], 143[k] | 95[l], 106[m], 118[o] | 111[u], 112[v], 95[w], 97[x], |
| $c_{13}$ | 58 | 50 | 86[a], 53[b], 94[c, j], 68[d], 64[e], 112[g], 174[h], 92[i], 144[k] | 54[l], 60[m] | 52[u], 55.3[v, x], |
| $c_{33}$ | 565 | 537 | 513[a], 566[b], 534[c, j], 498[d], 544[e], 453[g], 326[h], 510[i], 430[k] | 534[l], 566[m] | 553[u], 521.4[v, x], 565[w], |
| $c_{44}$ | 168 | 168 | 111[a], 192[b], 176[c, j], 167[d, i], 160[e], 142[g], 38[h], 129[k] | 161[l], 165[m], 238[o] | 163[u], 137[v], 169[w], 148.4[x], |
| $c_{66}$ | 196 | 189 | 129[a], 213[b], 192[c, j], 184[d], 181[e], 154[g], 39[h], 182[i], 132[k] | 193[l], 200[m] | 195[u], 203[w], 190.6[x], |
| $B'$ | (3.79) | (3.79) | (3.95)[a], (3.82)[d], (4.07)[e], (0.61)[g], (4.06)[h], | (3.84)[l], (3.77)[m], 3.2[n] | |

(4.01)[i], (6.23)[k]

**Table 5**. The dielectric parameters (dielectric constants ($\varepsilon_\infty$) and Born effective charges ($Z_B$)).

| | Present LDA | Present PBE | Other DFT | Experiment |
| --- | --- | --- | --- | --- |
| $\varepsilon_\infty(3C\text{-SiC})$ | 6.93 | 6.99 | 6.97[a], 6.88[b] | 6.52[c] |
| $\varepsilon_{\infty,a}(2H\text{-SiC})$ | 6.90 | 6.94 | 6.89[a] | |
| $\varepsilon_{\infty,c}(2H\text{-SiC})$ | 7.26 | 7.31 | 7.27[a] | |
| $\varepsilon_{\infty,a}(4H\text{-SiC})$ | 6.97 | 7.01 | 6.96[a] | |
| $\varepsilon_{\infty,c}(4H\text{-SiC})$ | 7.27 | 7.31 | 7.17[a] | |
| $\varepsilon_{\infty,a}(6H\text{-SiC})$ | 6.97 | 7.02 | | |
| $\varepsilon_{\infty,c}(6H\text{-SiC})$ | 7.22 | 7.26 | | |
| $Z_B(3C\text{-SiC})$ | 2.71 | 2.69 | 2.72[a], 2.71[b] | 2.69[d] |
| $Z_{B,a}(2H\text{-SiC})$ | 2.68 | 2.65 | 2.62[a] | |
| $Z_{B,c}(2H\text{-SiC})$ | 2.88 | 2.85 | 2.81[a] | |
| $Z_{B,a}(4H\text{-SiC})$ | 2.69* | 2.67* | 2.64[a] | |
| $Z_{B,c}(4H\text{-SiC})$ | 2.79* | 2.77* | 2.84[a] | |
| $Z_{B,a}(6H\text{-SiC})$ | 2.70** | 2.68** | | |
| $Z_{B,c}(6H\text{-SiC})$ | 2.76** | 2.74** | | |

a: LDA calculation [14]; b: LDA calculation [16]; c: experimental data present in Ref. [14]; d: experiment [127]; *: averaged two atoms; **: averaged three atoms

**Fig. 1.** (Color)
Comparison of the values predicted by DP models with the corresponding training datasets by *ab initio* calculations (DFT values). (a) energies of DPlda (eV/atom), (b) forces of DPlda (eV/Angstrom), (c) virial energies of DPlda (eV/atom), (d) energies of DPpbe (eV/atom), (e) forces of DPpbe (eV/Angstrom) and (f) virial energies of DPpbe (eV/atom). The solid lines (*y=x*) represent the identity function used to guide the eyes.

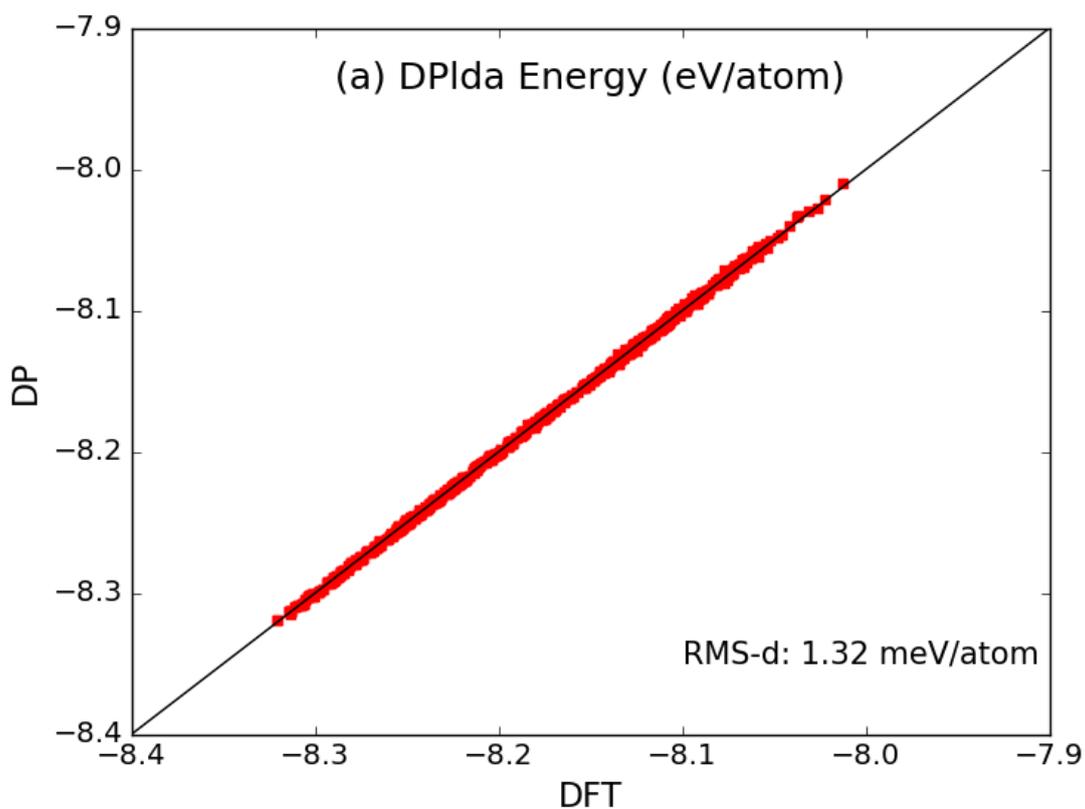

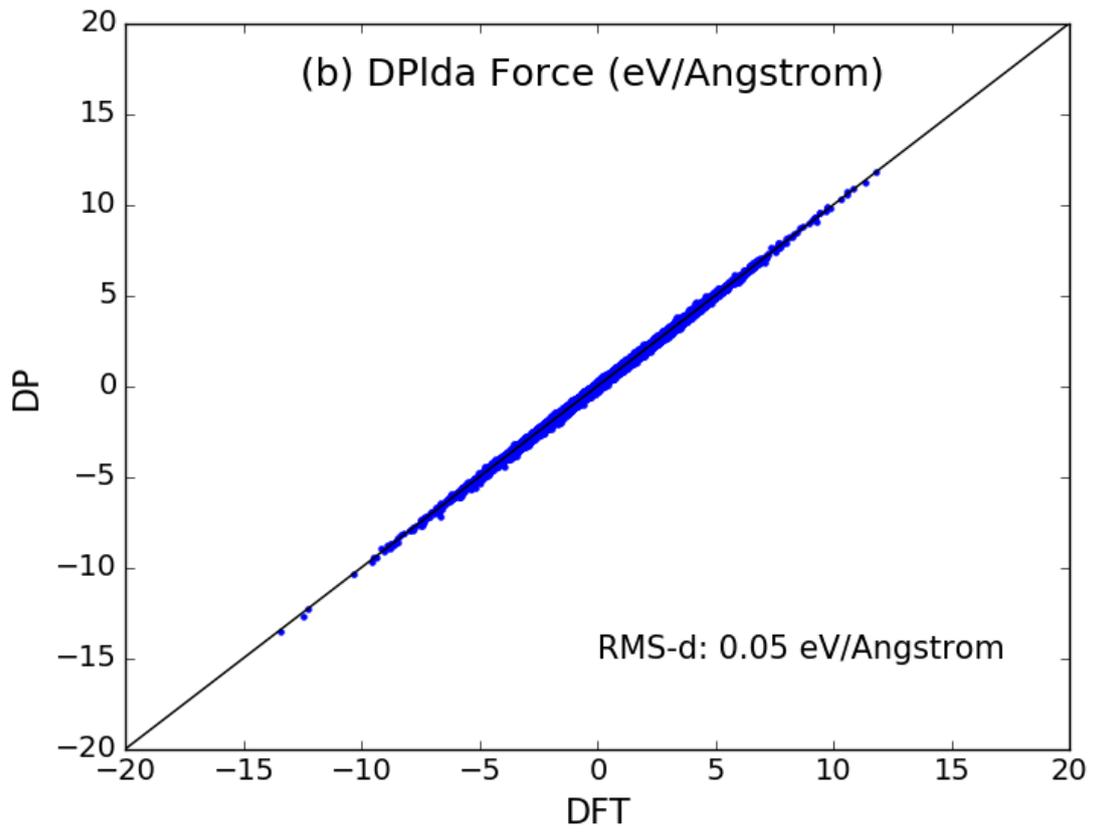

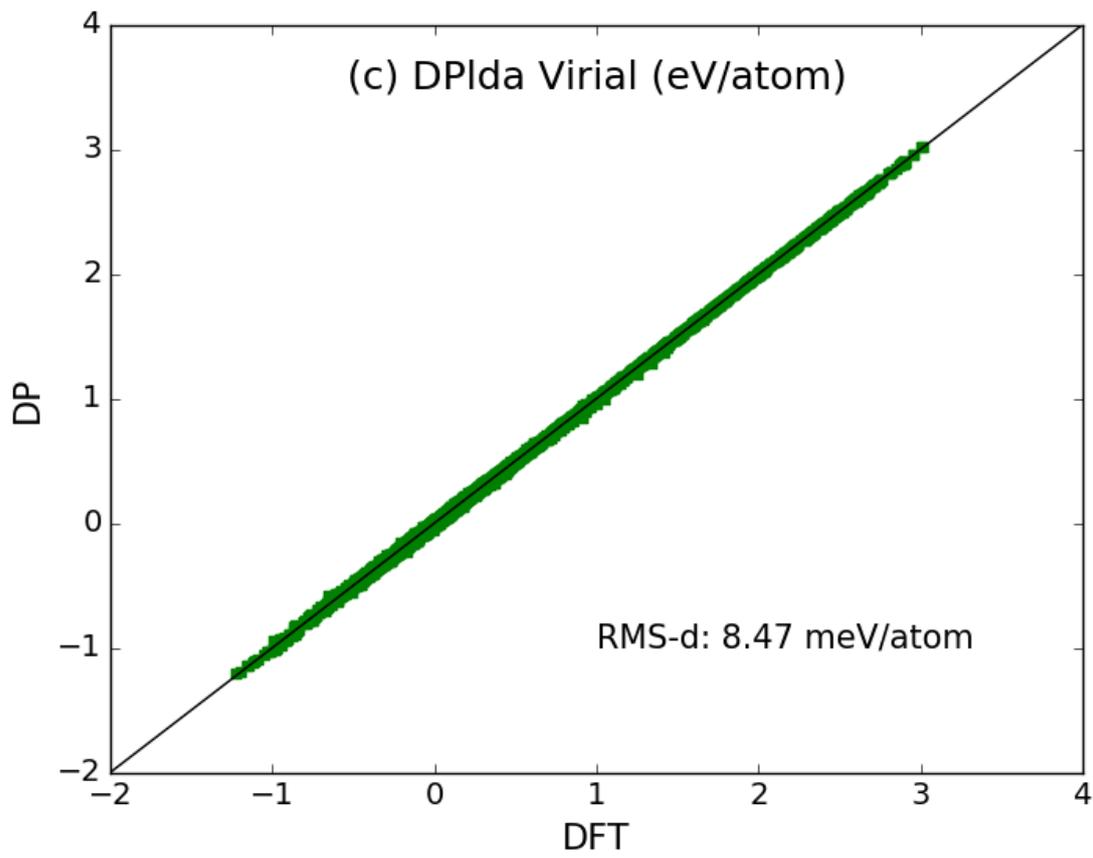

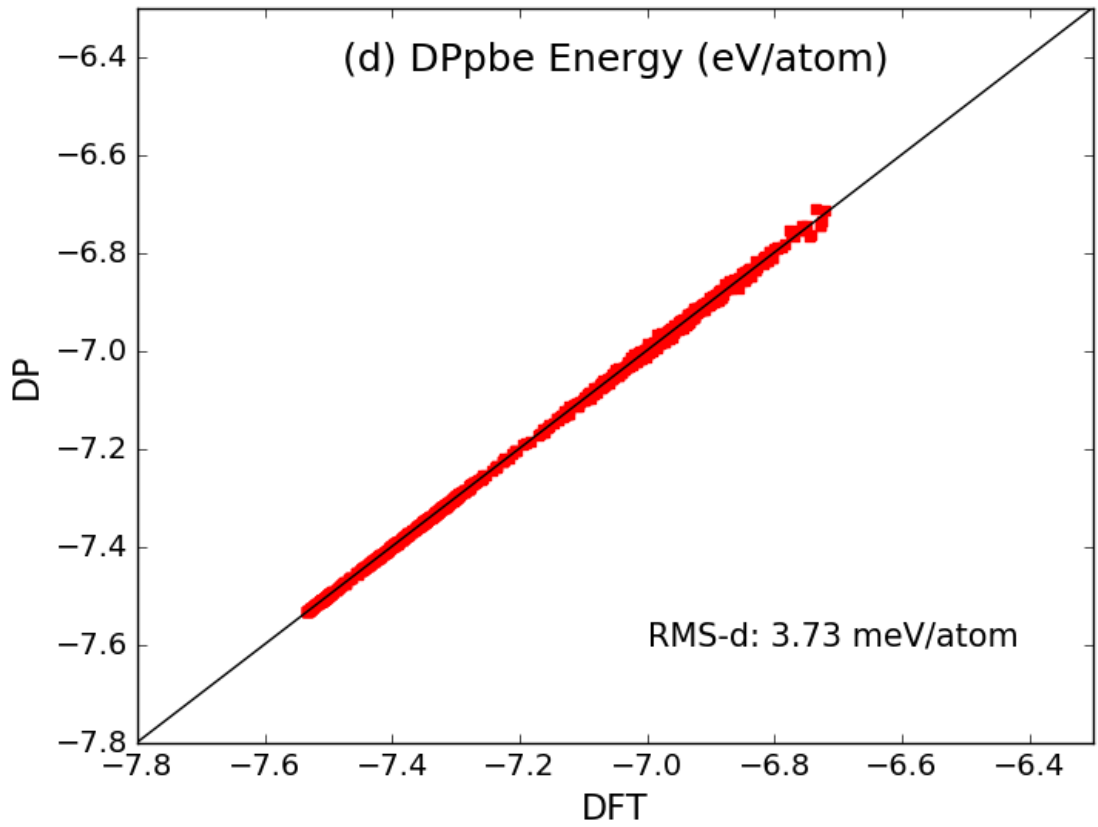

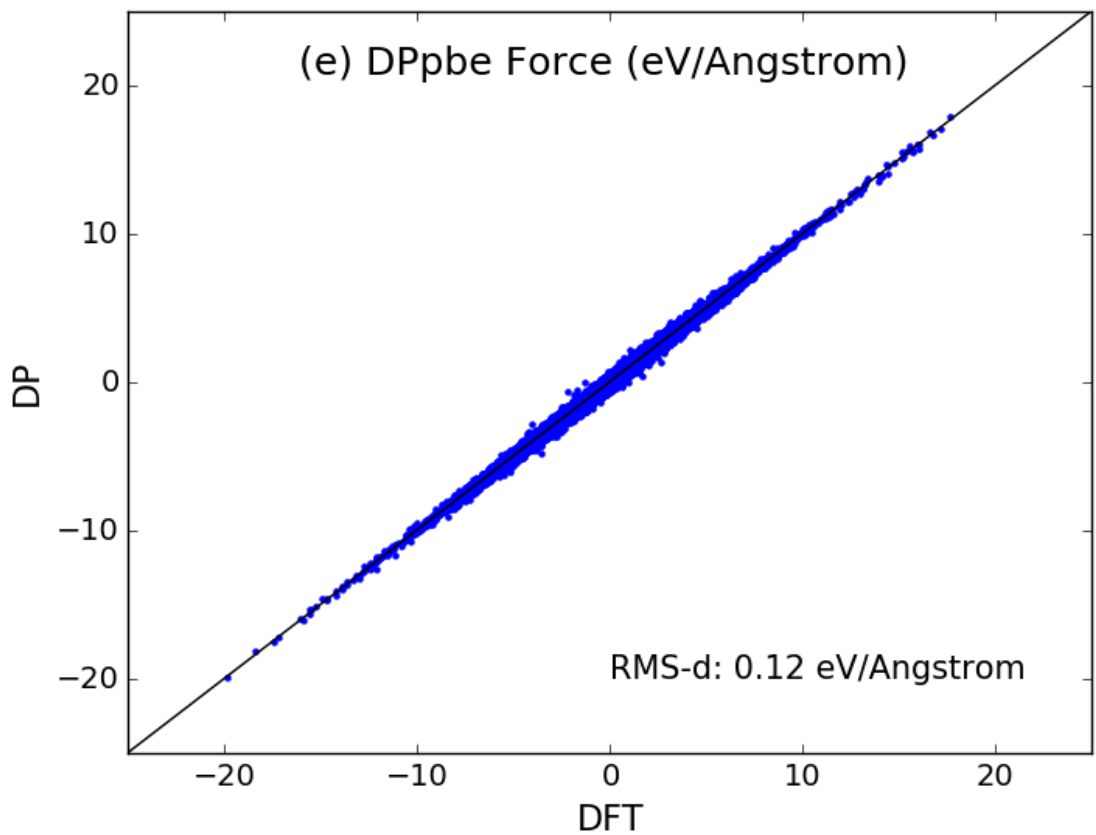

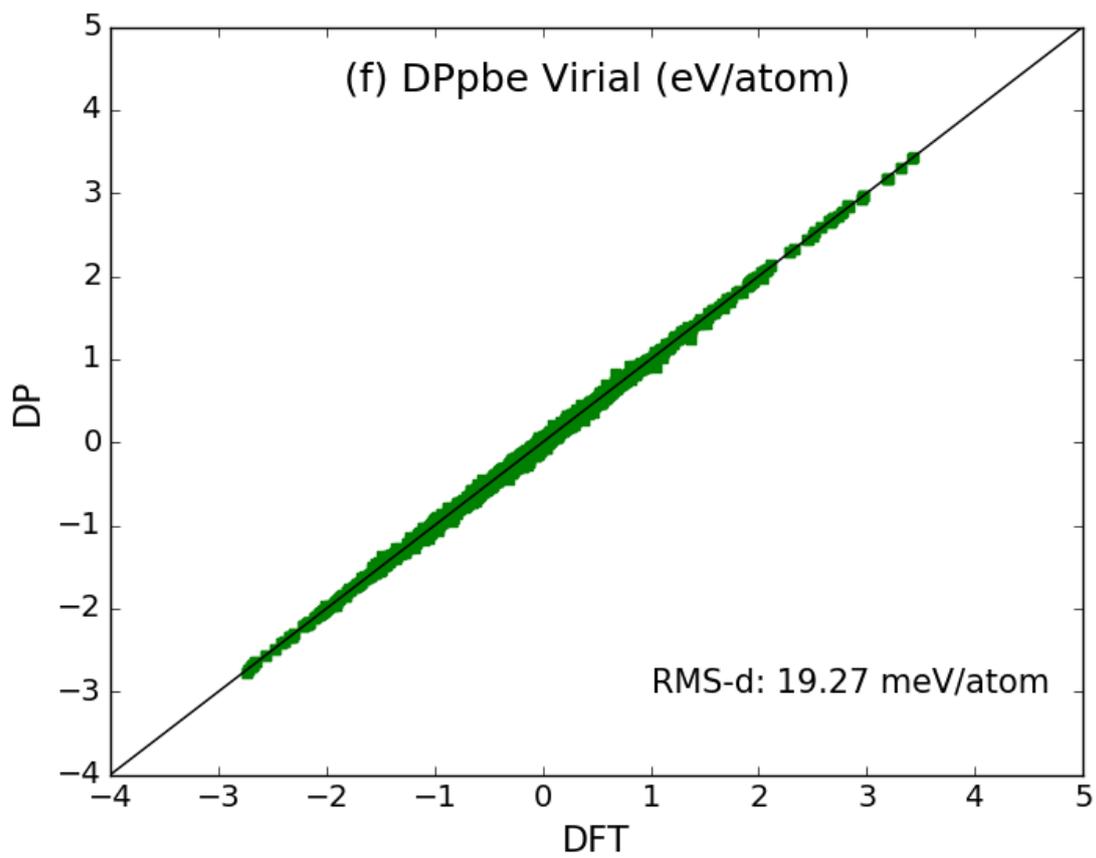

Fig. 2 (Color)

(a) The relationship between cohesive energy per atom ($E_c$, eV/atom) and volume per atom ($V$, Angstrom$^3$/atom) of 3$C$-SiC calculated by DFT methods (LDA and PBE), DP-IAPs (DPlda and DPpbe) and E-IAPs (DD98 [47], EA05 [49], GW02 [48], LB10 [51], HG95 [44], KE14 [52], T89 [39], T90 [41], T94 [42] and VK07[50]); (b) The $E_c$-$V$ relationships of 3$C$-SiC with DP-IAPs are fitted by Murnaghan equation of state (Eq. (2)) [100].

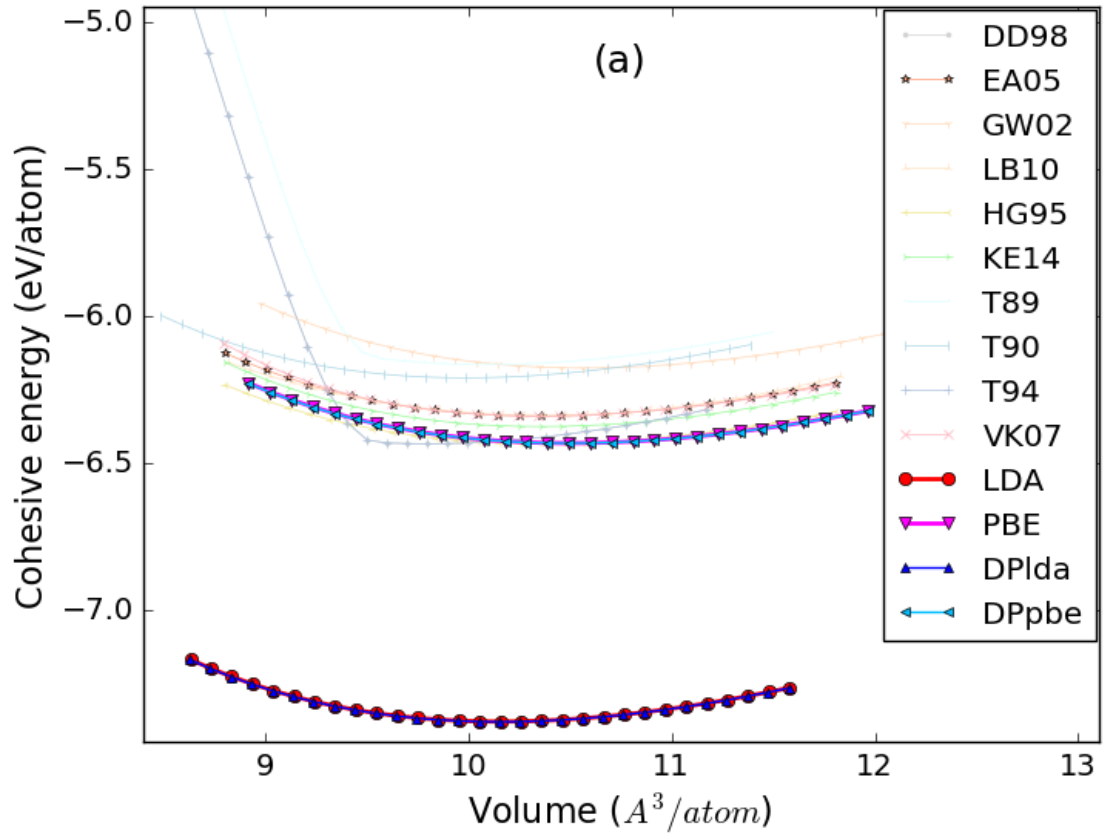

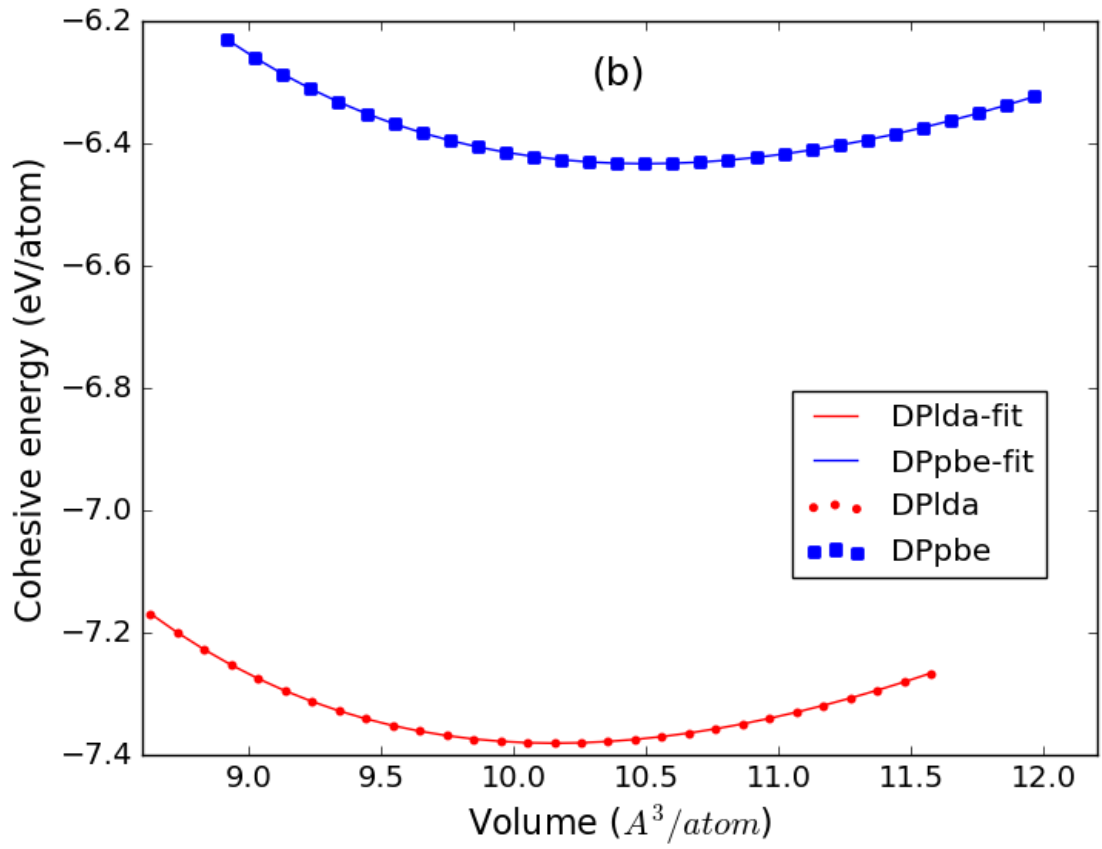

Fig. 3 (Color)

Phonon dispersion as given by two DP-IAPs (DPlda and DPpbe), compared with DFT data (LDA and PBE), experimental data (Exp. Feldman [132], Exp. Serrano [124], Exp. Widulle [133] and Exp. Nakashima [136]) and the data of E-IAPs (DD98 [47], EA05 [49], GW02 [48], LB10 [51], HG95 [44], KE14 [52], T89 [39], T90 [41], T94 [42] and VK07[50]). (a) 3*C*-SiC, (b) 2*H*-SiC, (c) 4*H*-SiC, (d) 6*H*-SiC.

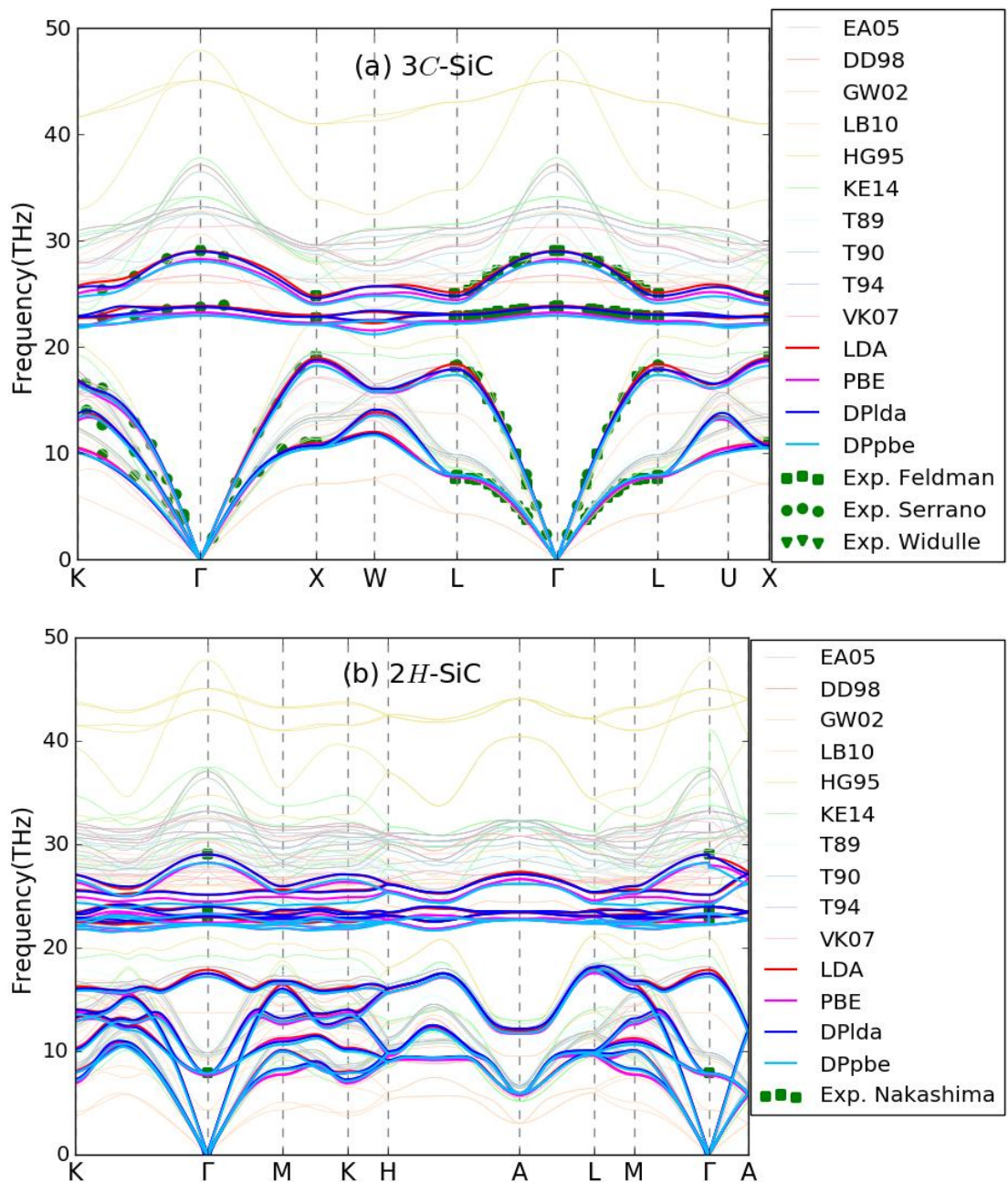

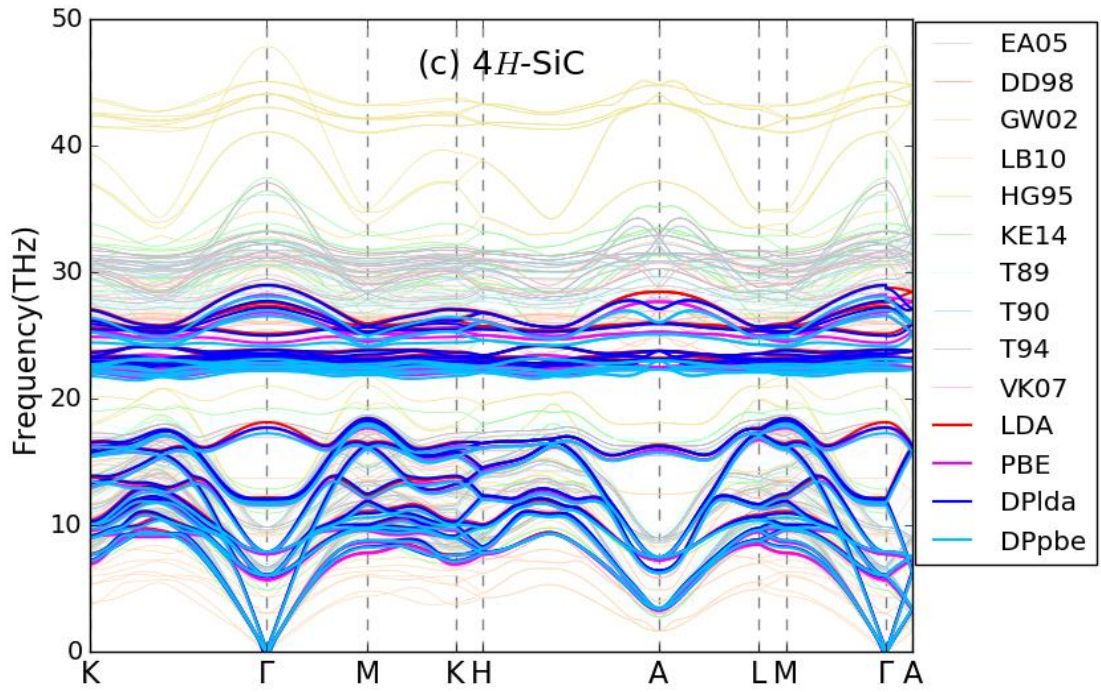

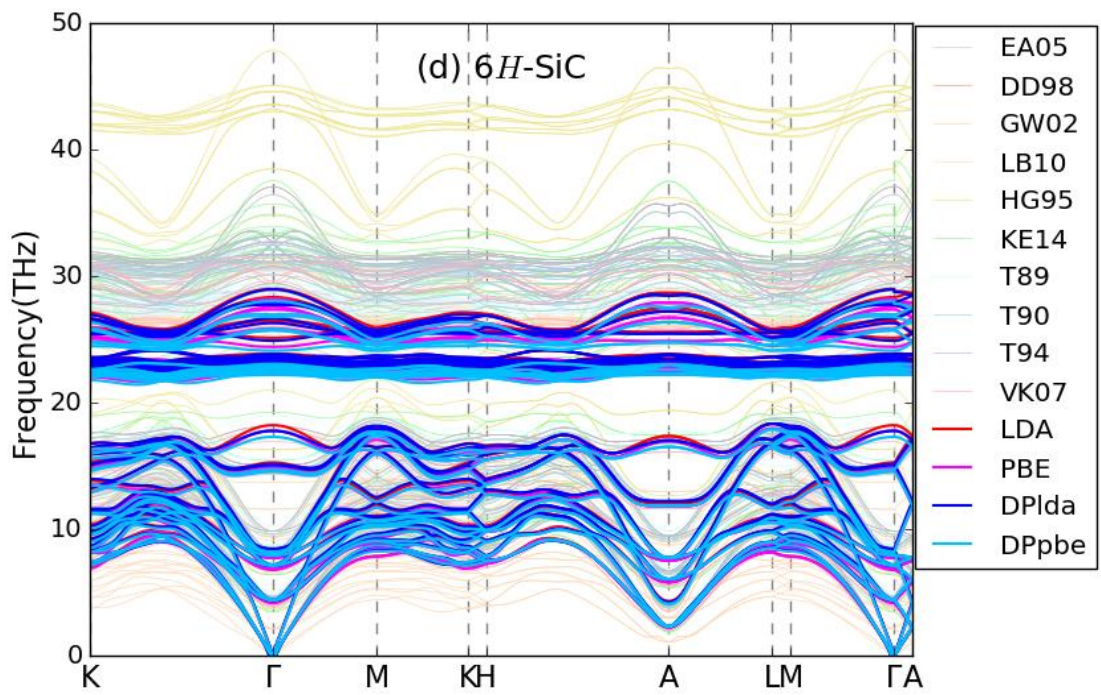

Fig. 4 (Color)

Density of states (DOS) of phonon as given by two DP-IAPs (DPlda and DPpbe), compared with DFT data (LDA and PBE) and the data of E-IAPs (DD98 [47], EA05 [49], GW02 [48], LB10 [51], HG95 [44], KE14 [52], T89 [39], T90 [41], T94 [42] and VK07[50]). (a) 3$C$-SiC, (b) 2$H$-SiC, (c) 4$H$-SiC, (d) 6$H$-SiC.

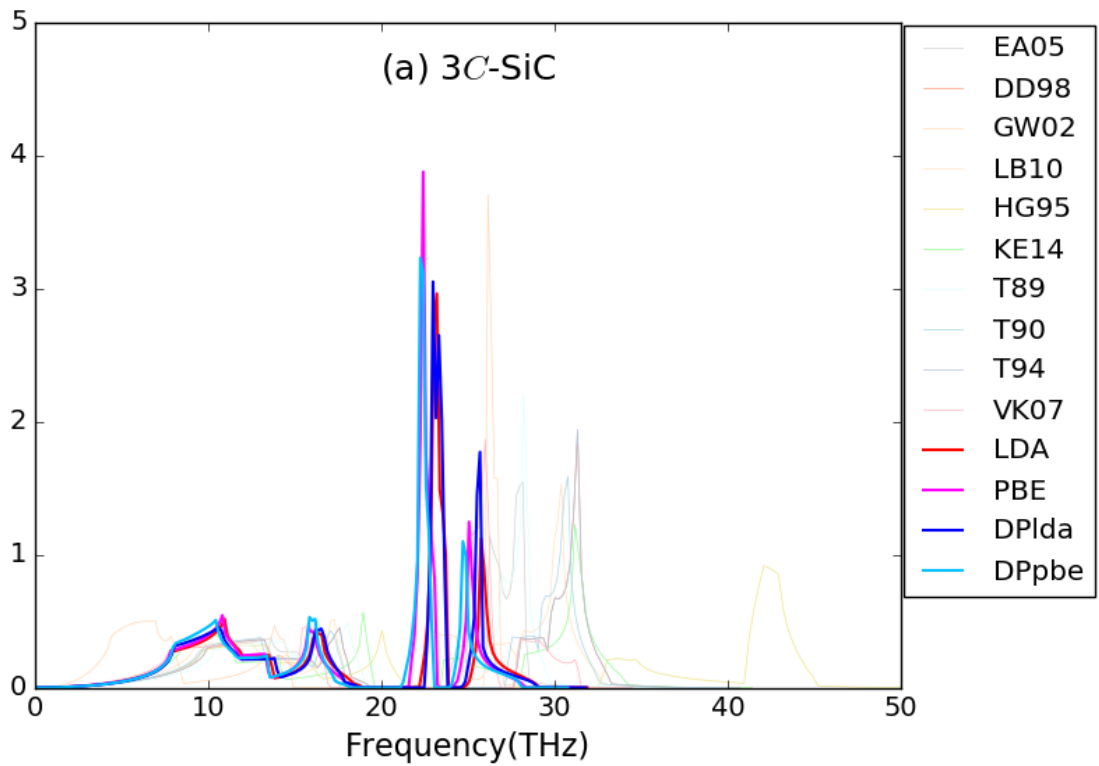

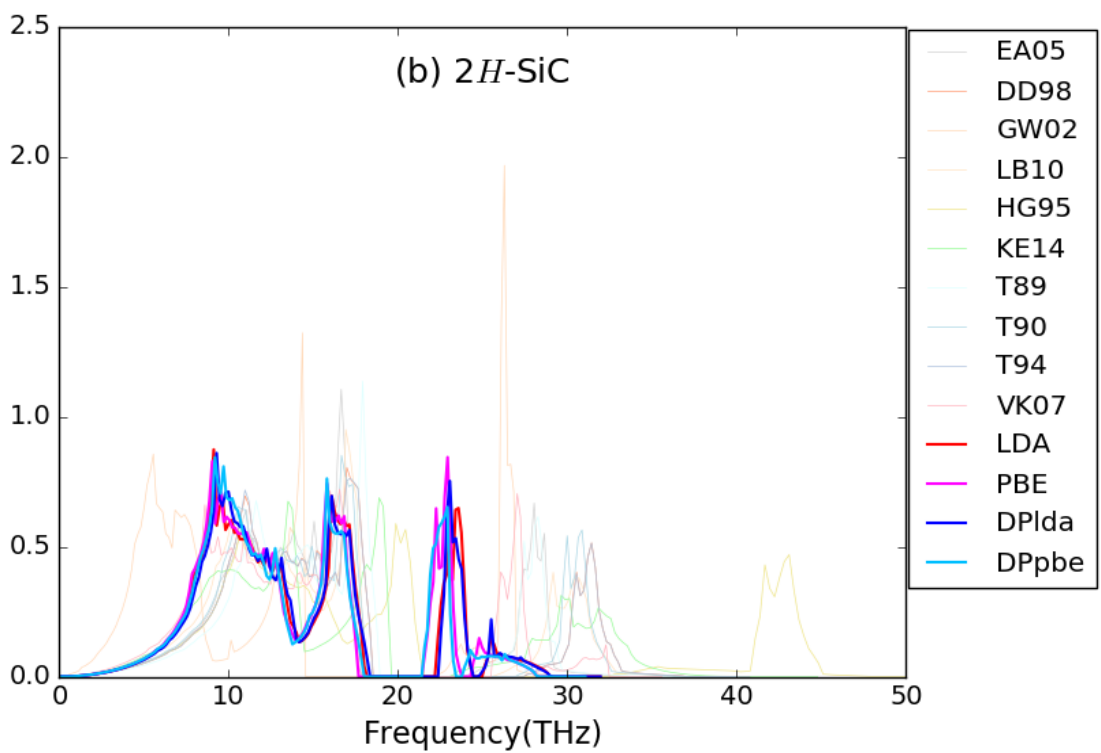

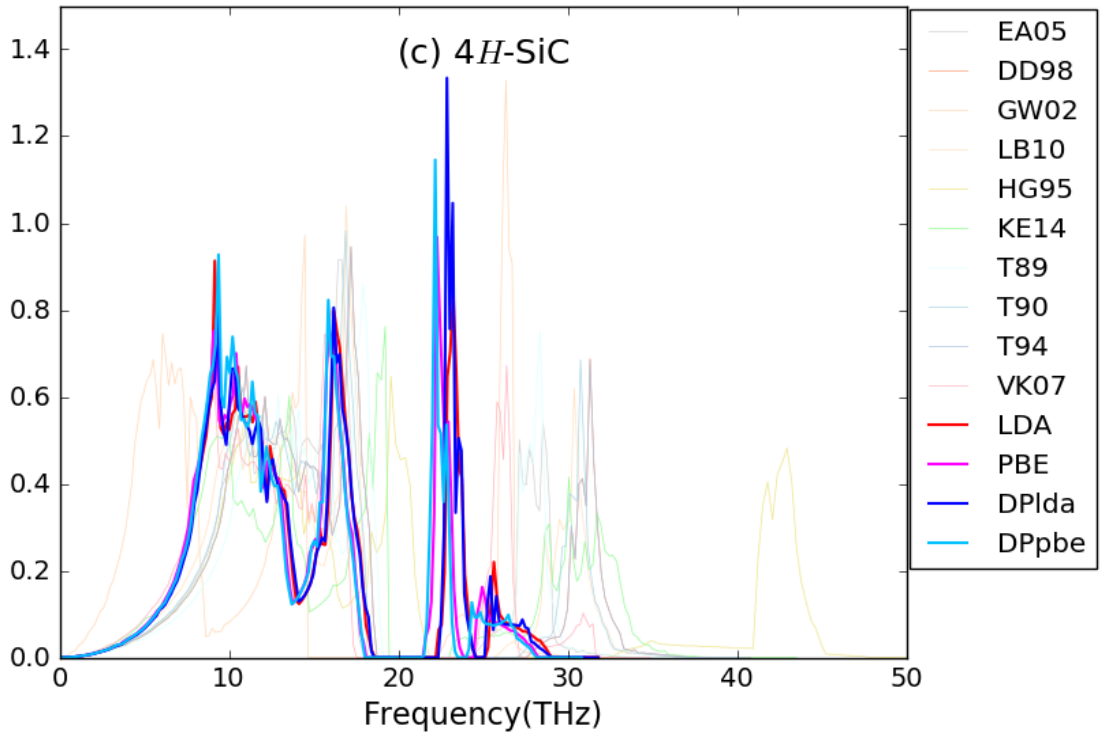

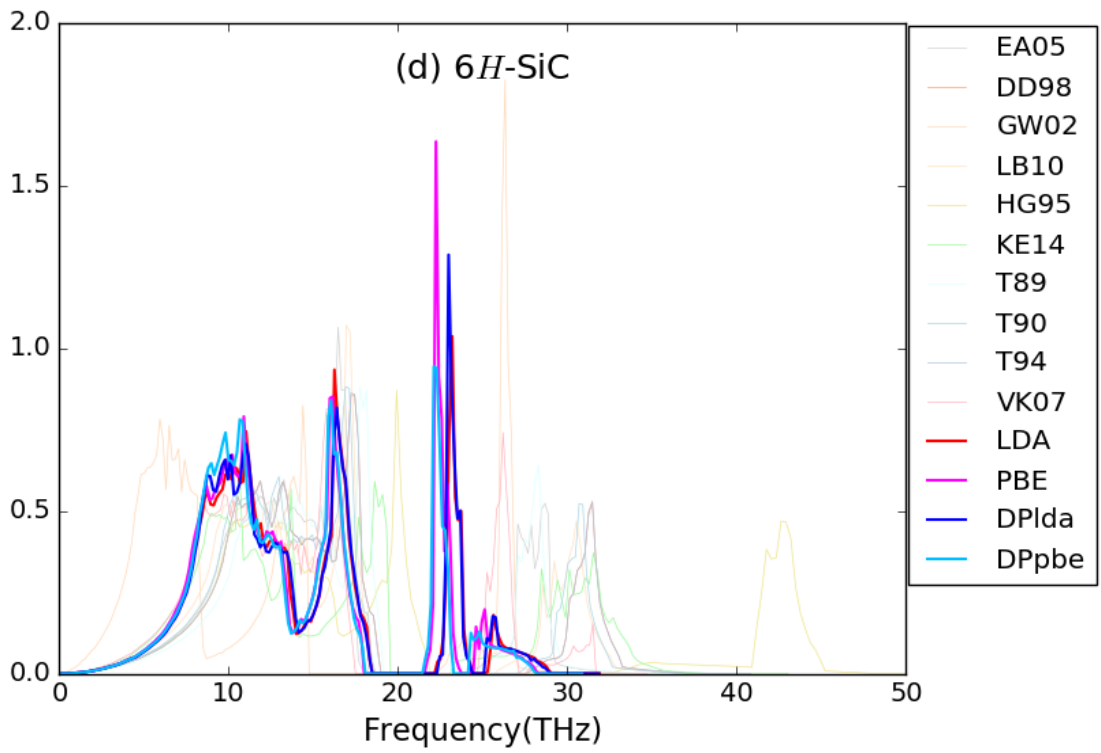

Fig. 5 (Color)

Entropy ($S$, J/K/mol), Helmholtz free energy ($F$, kJ/mol) and heat capacity at constant volume ($C_V$, J/K/mol) as function of temperatures ($T$) given by two DP-IAPs (DPlda and DPpbe), compared with DFT data (LDA and PBE) and the data of E-IAPs (DD98 [47], EA05 [49], GW02 [48], LB10 [51], HG95 [44], KE14 [52], T89 [39], T90 [41], T94 [42] and VK07[50]). (a) 3$C$-SiC (two atoms in unit cell), Exp. $S$ denotes the experimental data from Ref. [140], (b) 2$H$-SiC (four atoms in unit cell), (c) 4$H$-SiC (eight atoms in unit cell), (d) 6$H$-SiC (twelve atoms in unit cell).

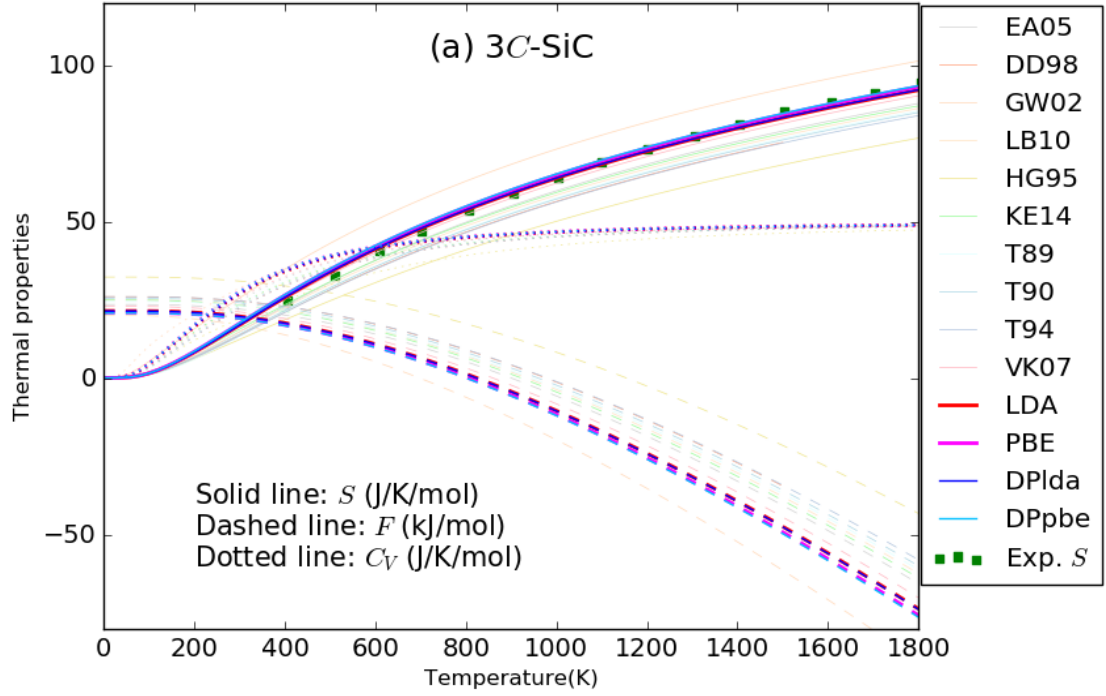

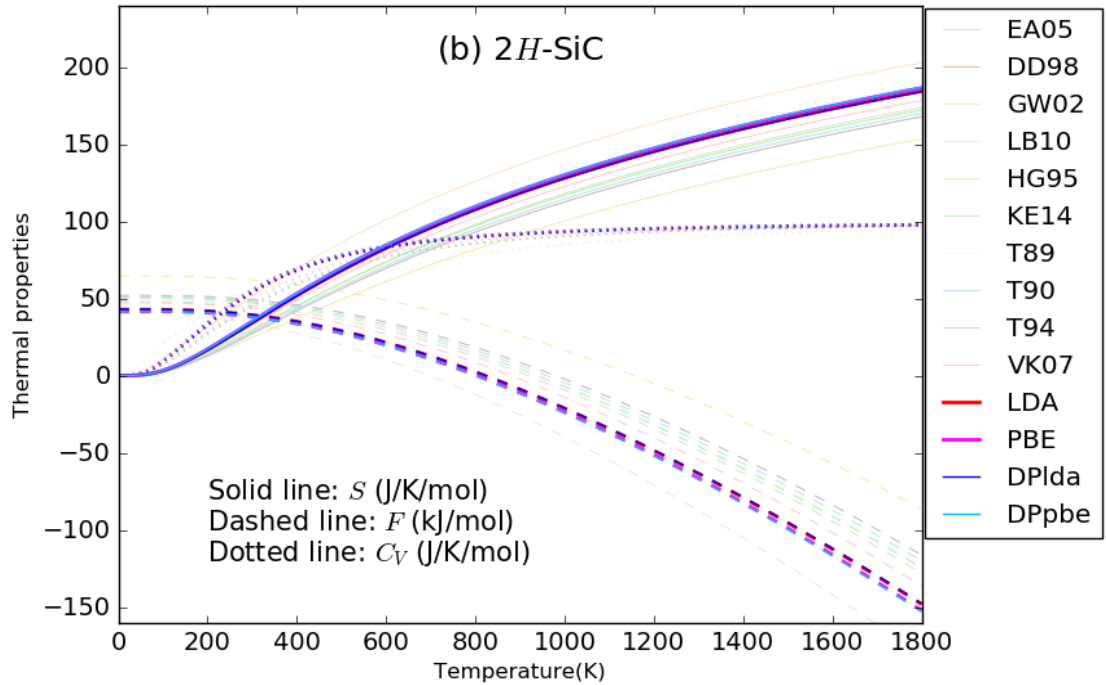

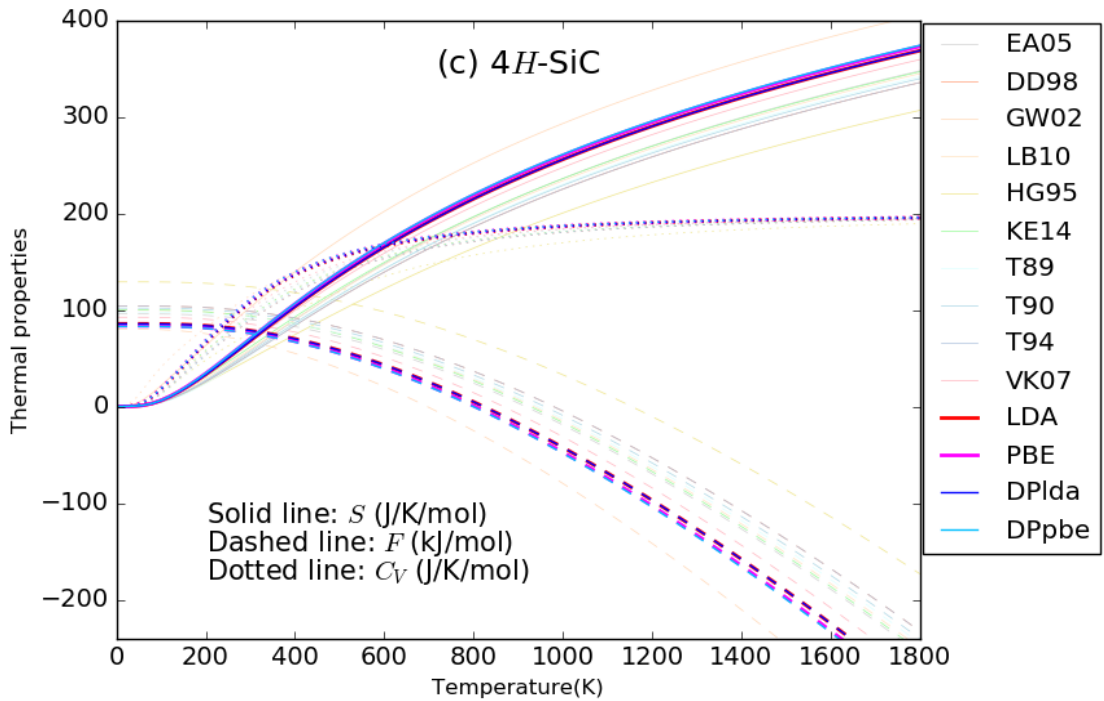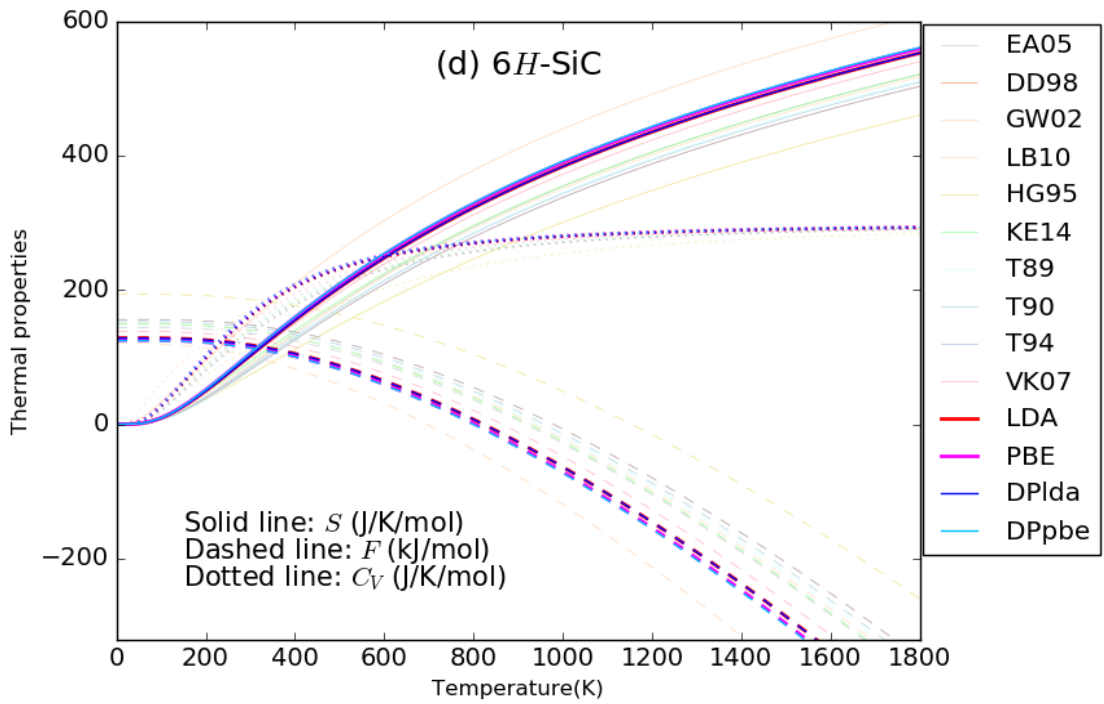

Fig. 6 (Color)

(a) Gibbs free energy ($G$, eV) of unit cell, (b) Volume ($V$, Angstrom$^3$) of unit cell, (c) bulk modulus ($B$, GPa), (d) the derivative of entropy with respect to volume (d$S$/d$V$, eV/K/Angstrom$^3$) and (e) Gruneisen parameter ($\gamma$) calculated by QHA method as functions of temperatures ($T$) given by two DP-IAPs (DPlda and DPpbe), compared with DFT data (LDA and PBE) and the data of E-IAPs (DD98 [47], EA05 [49], GW02 [48], LB10 [51], HG95 [44], KE14 [52], T89 [39], T90 [41], T94 [42] and VK07[50]) for 3$C$-SiC (two atoms in unit cell).

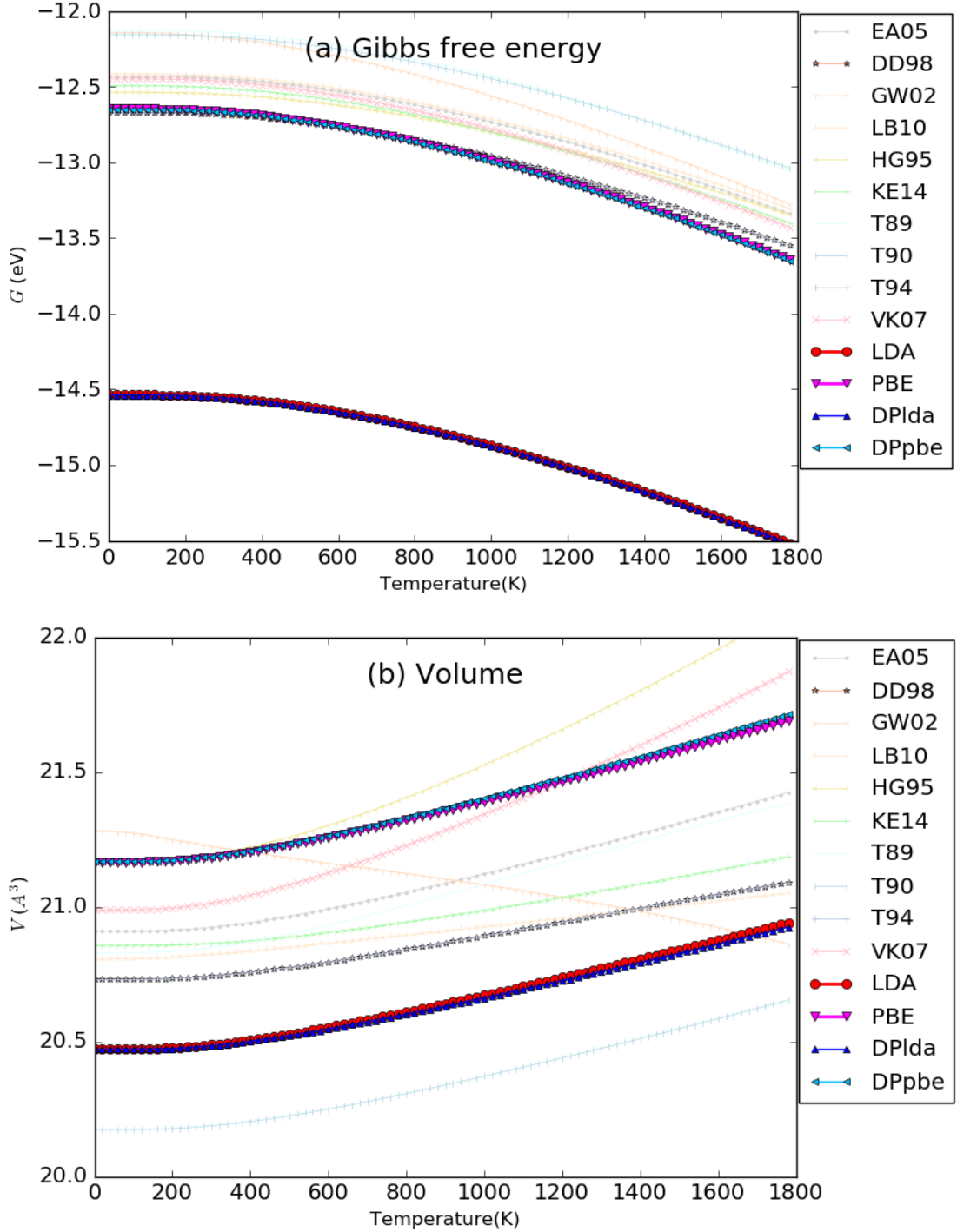

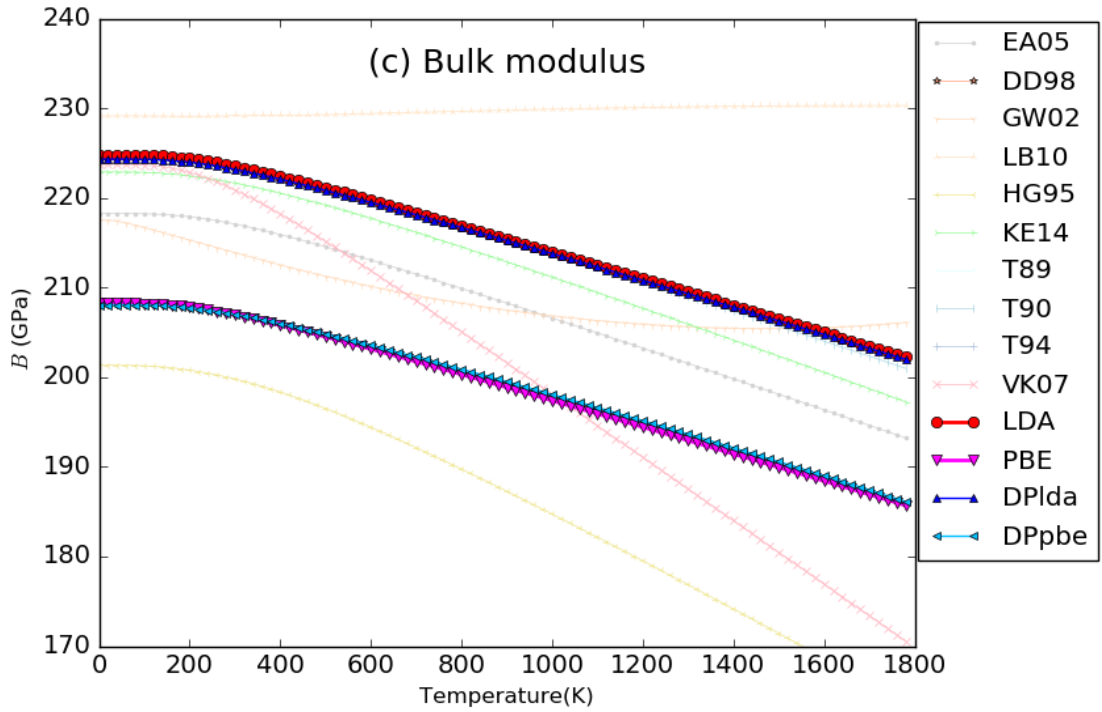

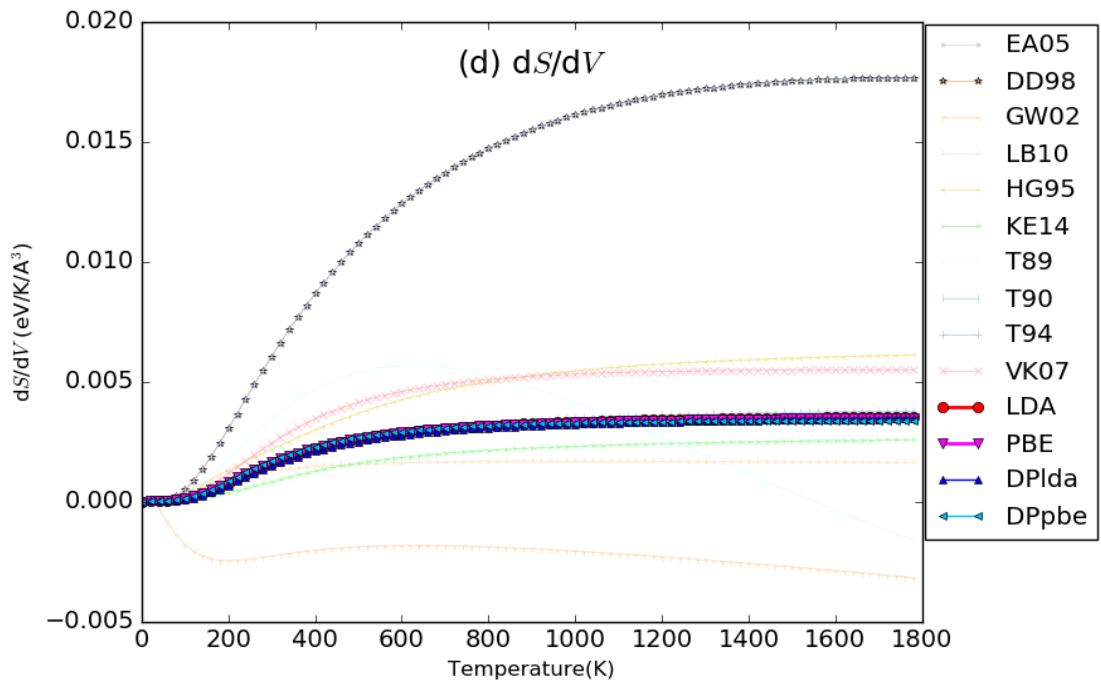

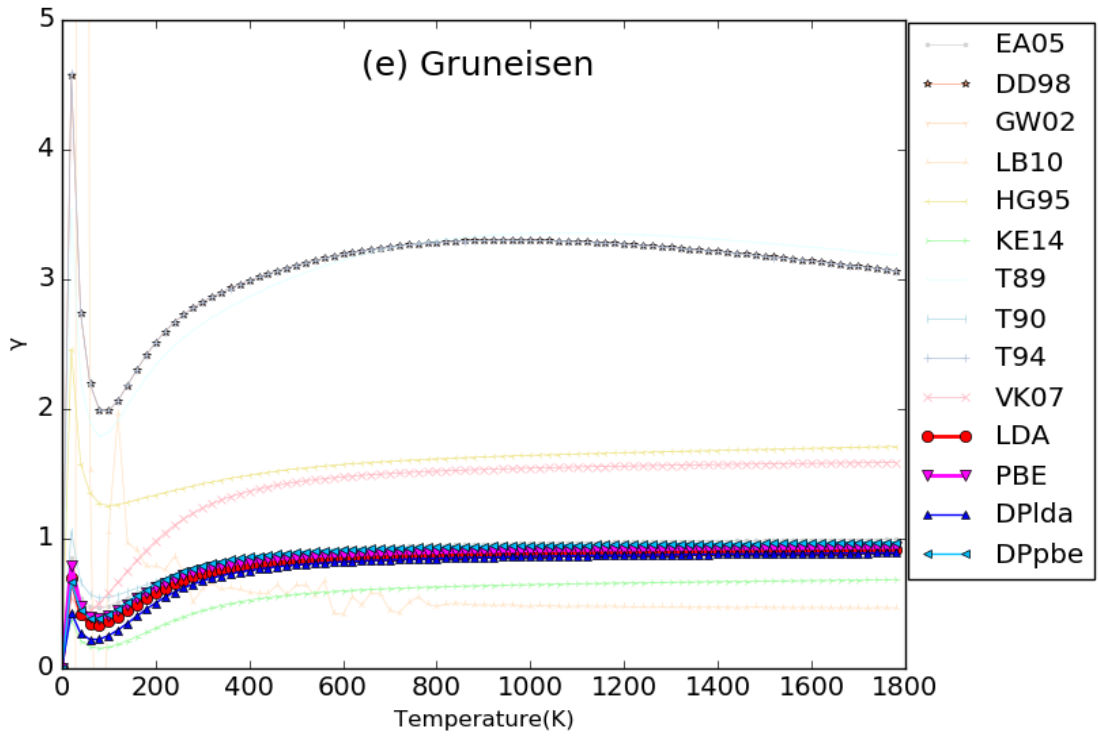

Fig. 7 (Color)

Relative bulk modulus ($R_B=B(T)/B(0)$) as function of temperatures ($T$) given by two DP-IAPs (DPlda and DPpbe), compared with DFT data (LDA and PBE), experimental data (Exp.1 [143,144], Exp.2 [141], Exp.3 [142], Exp.4 [147], Exp.5 [148], Exp.6 [146], Exp.7 [2] and Exp.8 [145]) originally compiled in Ref. [2], and the data of E-IAPs (DD98 [47], EA05 [49], GW02 [48], LB10 [51], HG95 [44], KE14 [52], T89 [39], T90 [41], T94 [42] and VK07[50]).

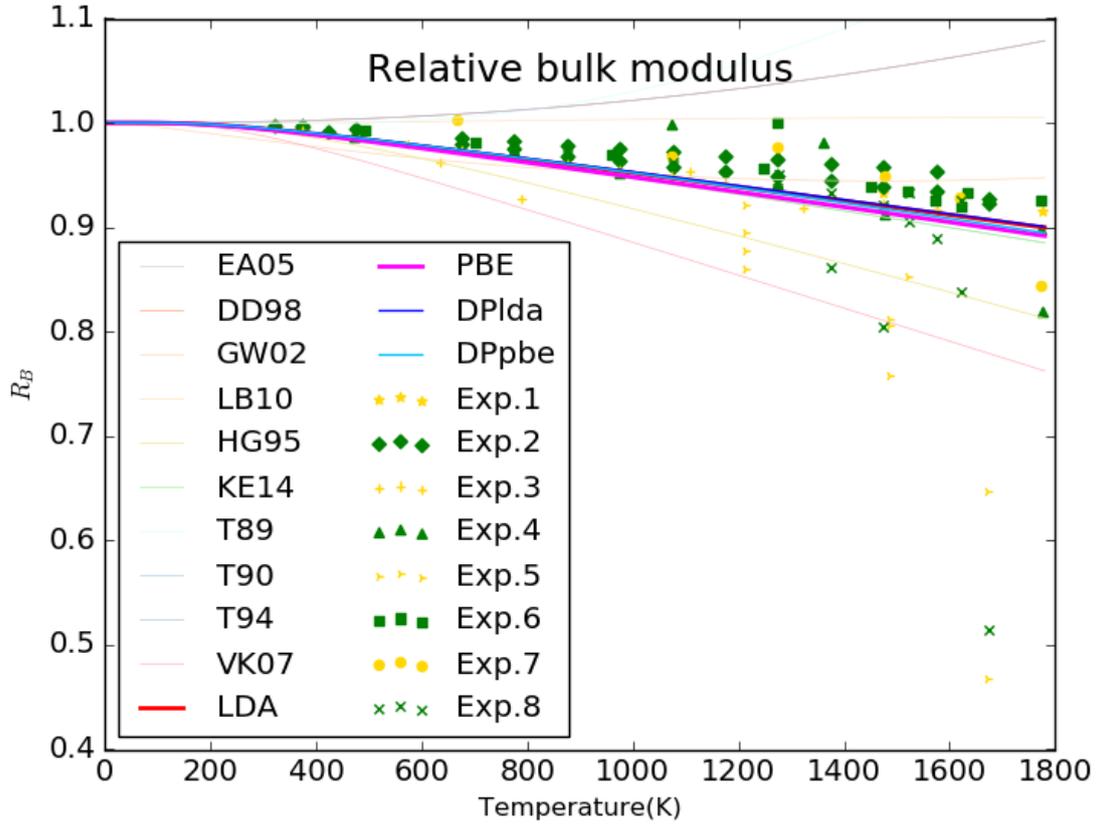

Fig. 8 (Color)

Heat capacity at constant pressure ($C_p$, J/K/mol) as function of temperatures ($T$) given by two DP-IAPs (DPlda and DPpbe), compared with DFT data (LDA and PBE), experimental data (Exp. Barin95 [140], Exp. Barin77 [150], Exp. Snead32 (Ref. 32 in [2]), Exp. Pickering [6], Exp. Collins [7], Exp. Kelley [8], Exp. Taylor [9] and Exp. Gurvich [149]) and the data of E-IAPs (DD98 [47], EA05 [49], GW02 [48], LB10 [51], HG95 [44], KE14 [52], T89 [39], T90 [41], T94 [42] and VK07[50]) for 3$C$-SiC (two atoms in unit cell).

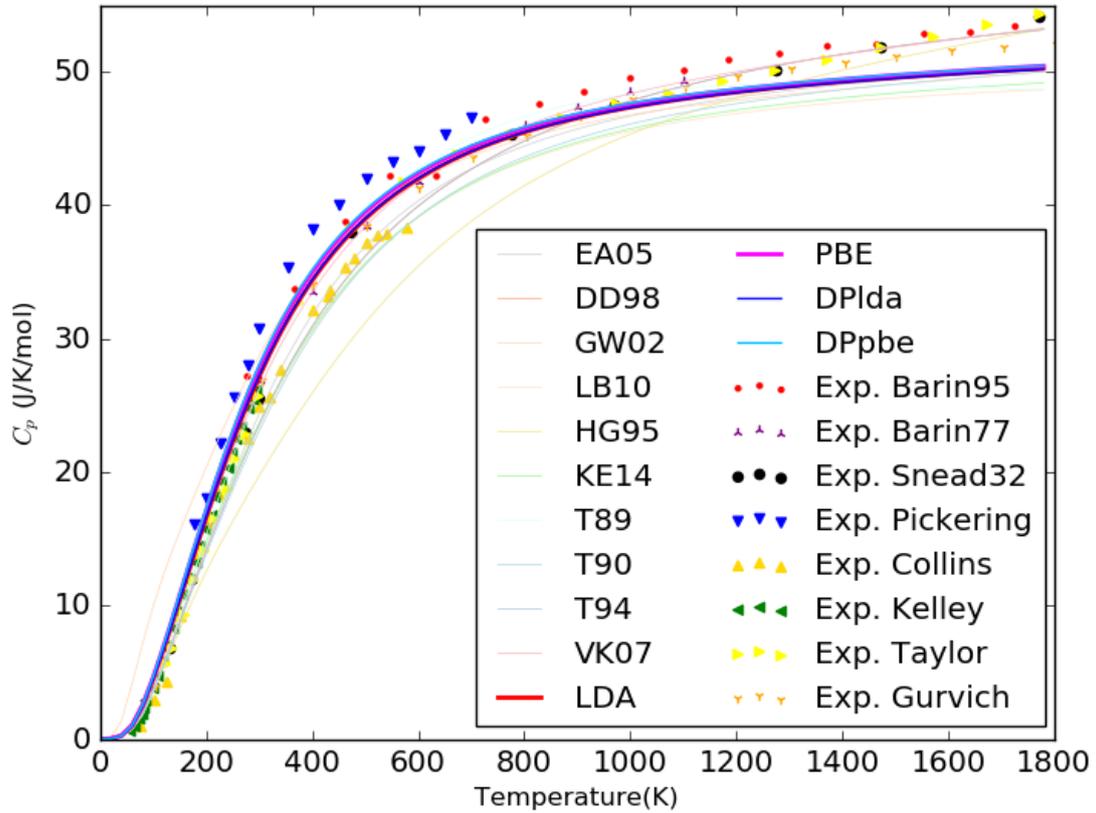

Fig. 9 (Color)

Coefficient of linear thermal expansion (CLTE) as function of temperatures ($T$) given by two DP-IAPs (DPlda and DPpbe), compared with DFT data (LDA, PBE and Karch's result [108]), experimental data (Exp. Li [10], Exp. Taylor60 [151], Exp. Pickering [6], Exp. Snead32 (Ref. 32 in [2]), Exp. Suzuki [11], Exp. Pojur [12] and Exp. Taylor93 [9]) and the data of E-IAPs (DD98 [47], EA05 [49], GW02 [48], LB10 [51], HG95 [44], KE14 [52], T89 [39], T90 [41], T94 [42] and VK07[50]) for 3$C$-SiC (two atoms in unit cell).

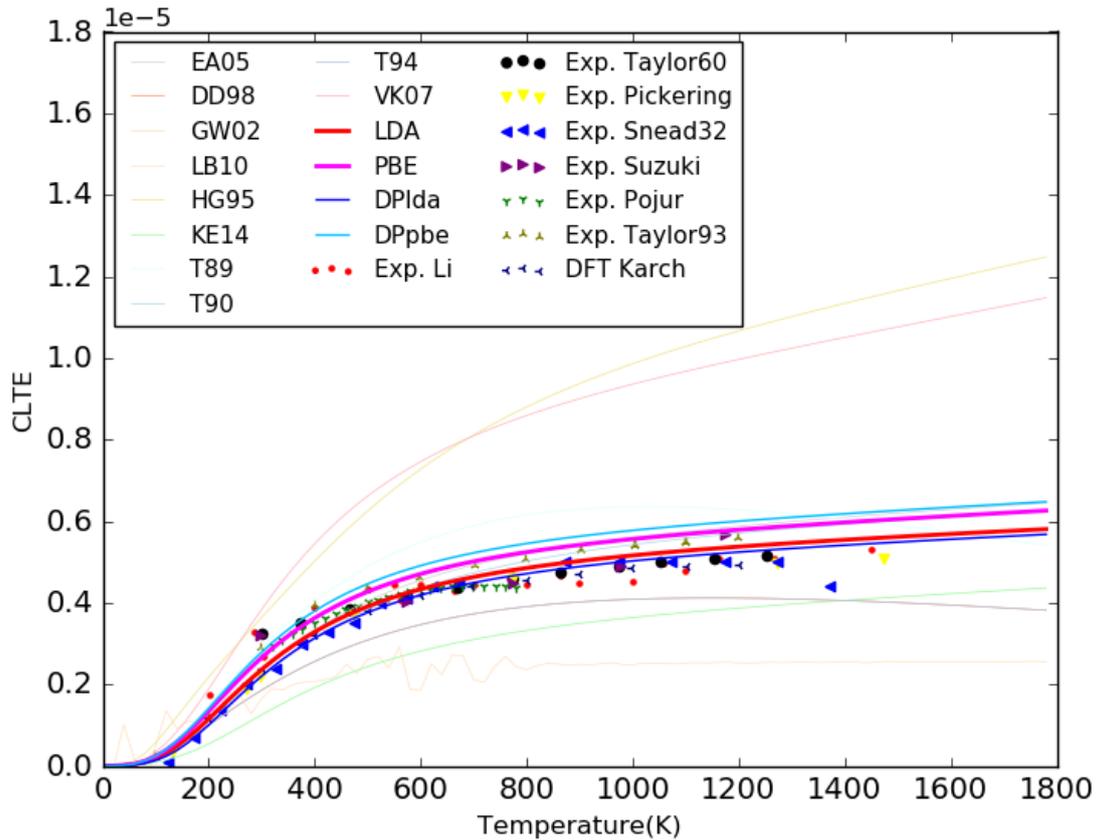

Fig. 10 (Color)

The BTE lattice thermal conductivity $\kappa_L$ of bulk 3$C$-SiC as function of temperatures ($T$) given by two DP-IAPs (DPlda and DPpbe), compared with DFT data (LDA and PBE), E-IAPs (EA05 [49] and T90 [41]) and experimental data (Taylor93 [9], Snead32 (Ref. 32 in [2]), Graebner98 [153], Slack64 [154], Morelli94 [155] and Senor96 [156]).

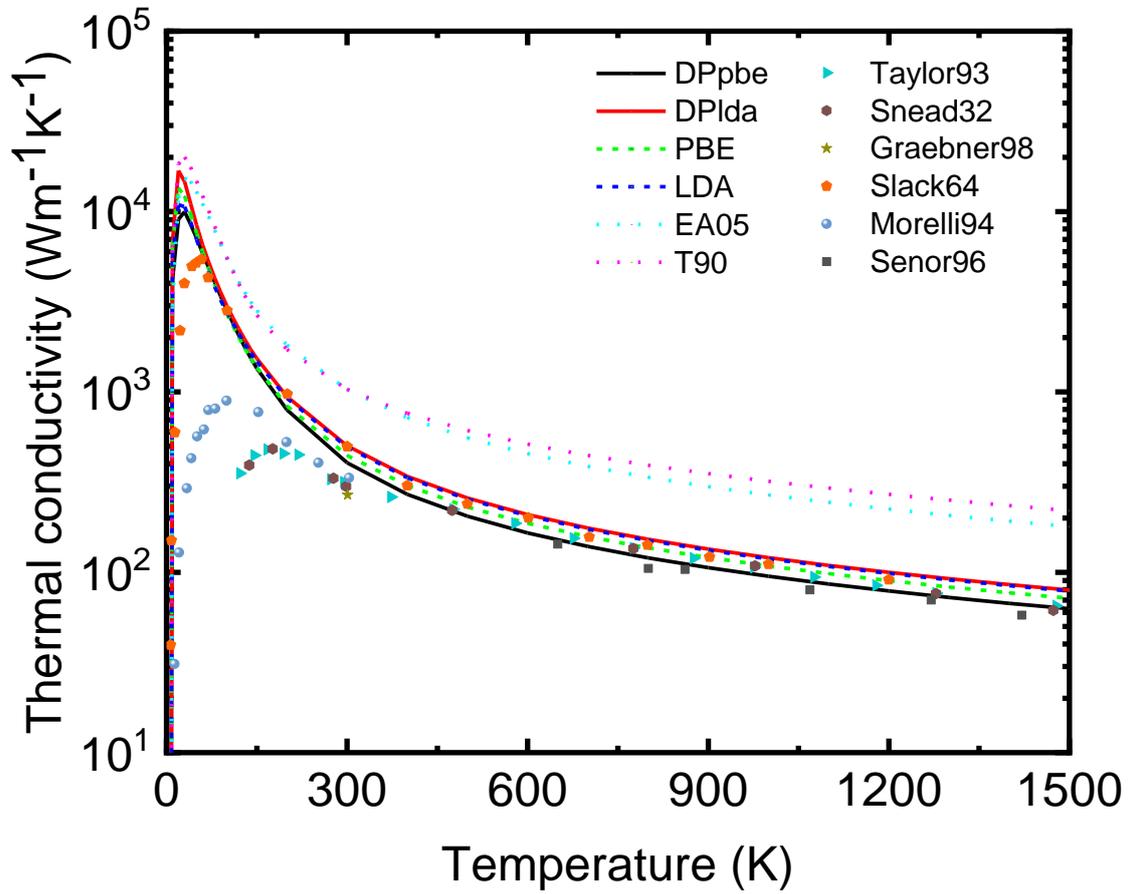

Fig. 11 (Color)

The phonon scattering rate $\Gamma$ from the BTE calculation of bulk 3*C*-SiC given by two DP-IAPs ((a) DPpbe and (b) DPlda), compared with DFT data ((c) PBE and (d) LDA) and E-IAPs data ((e) EA05 [49] and (f) T90 [41]) .

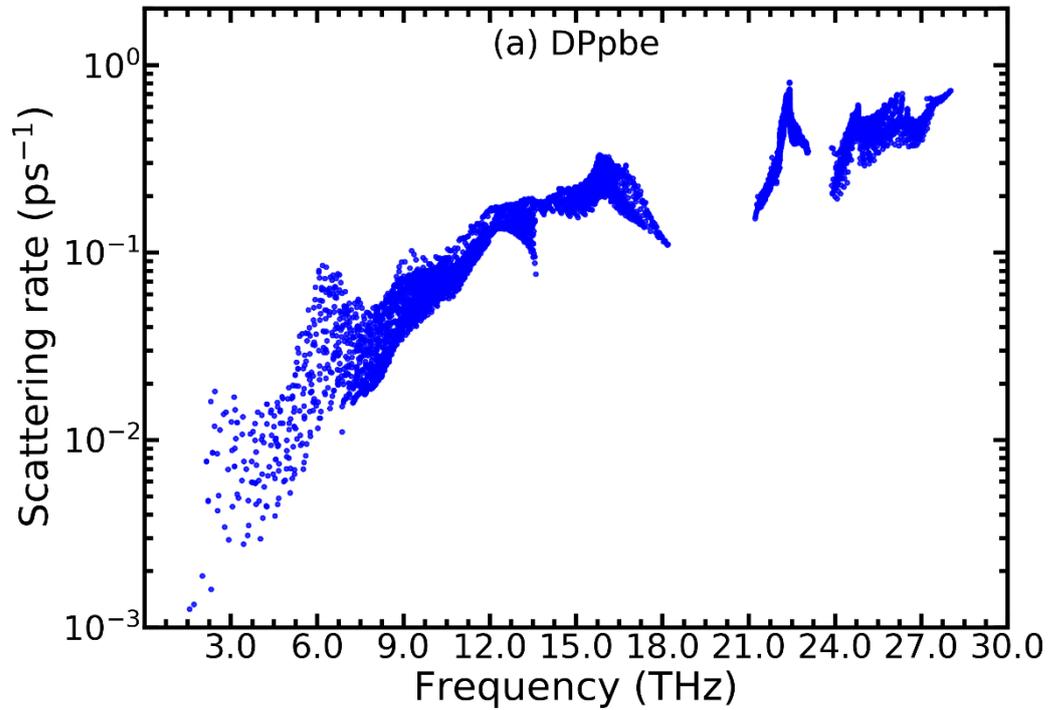

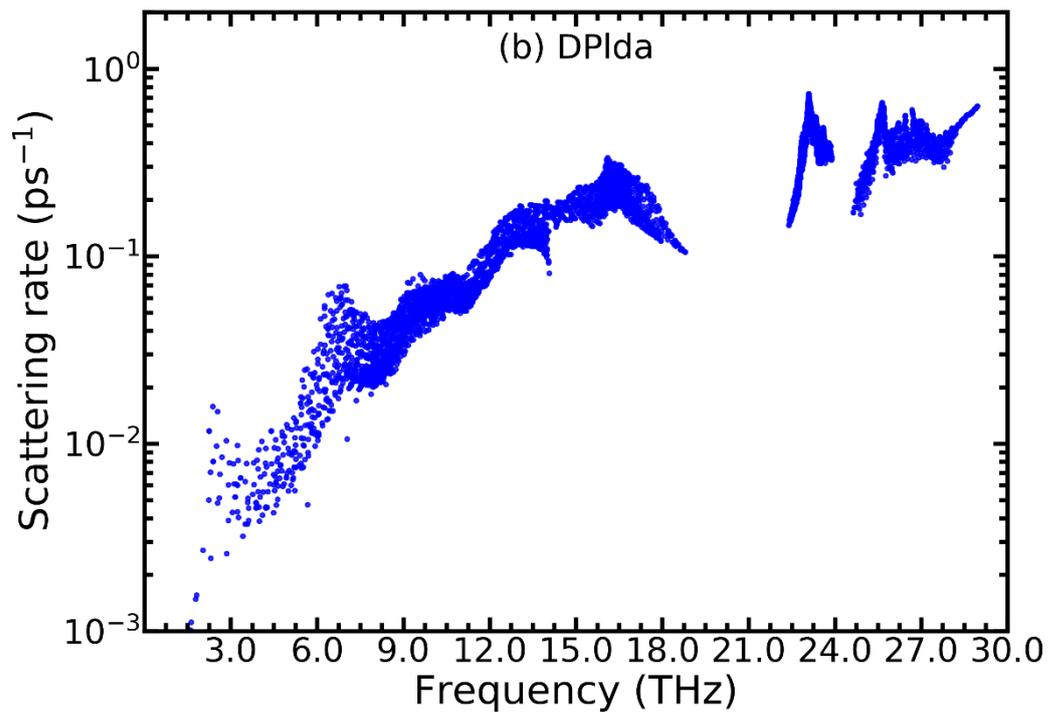

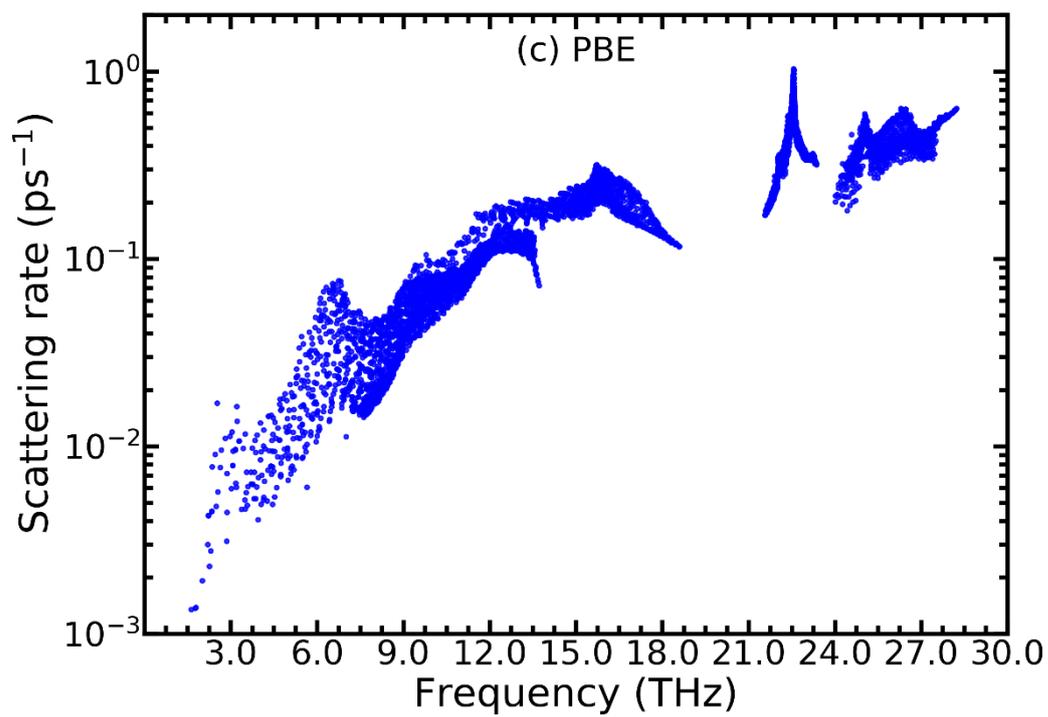
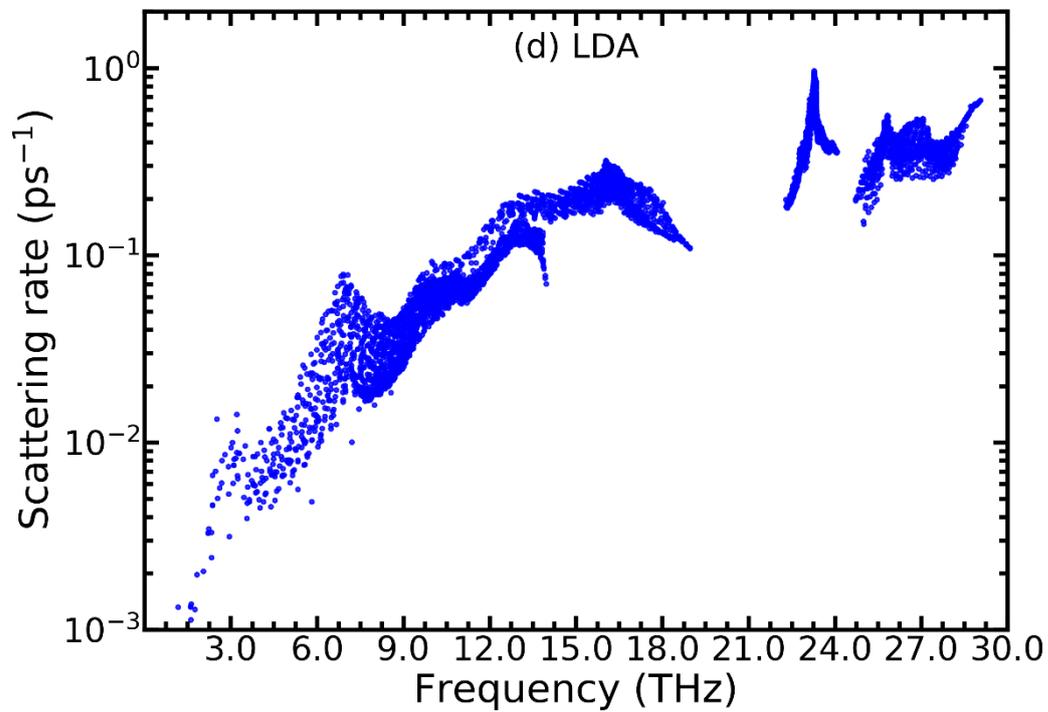

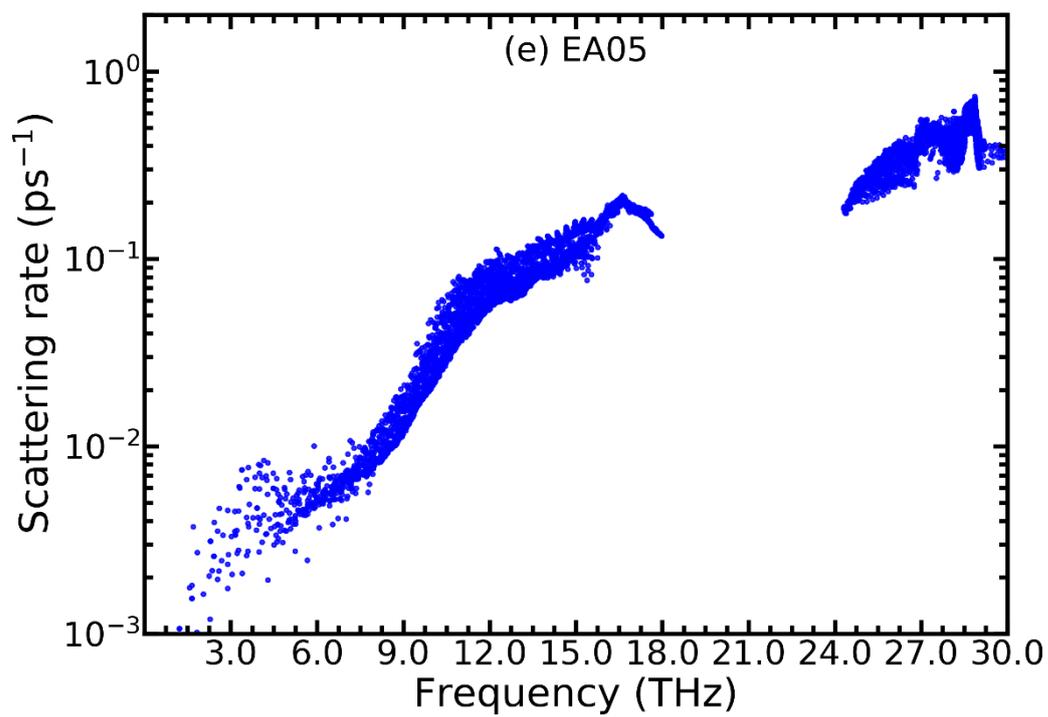

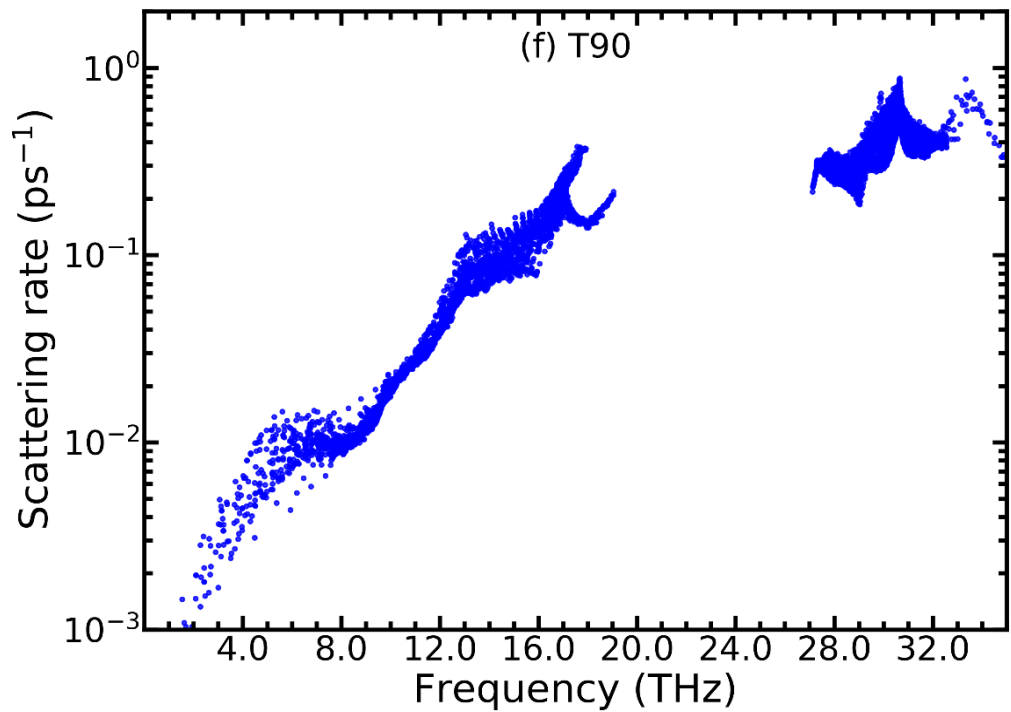

Fig. 12 (Color)
The mode Grüneisen parameter ($\gamma$) from the BTE calculation of bulk 3$C$-SiC given by two DP-IAPs ((a) DPpbe and (b) DPlda), compared with DFT data ((c) PBE and (d) LDA) and E-IAPs data ((e) EA05 [49] and (f) T90 [41]) .

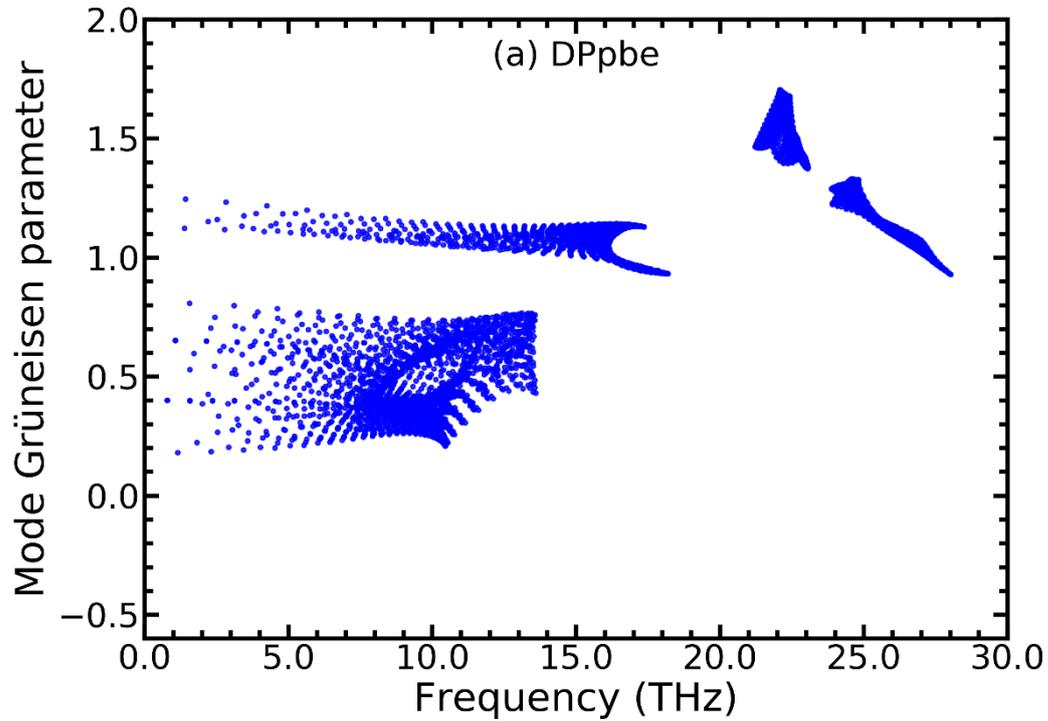

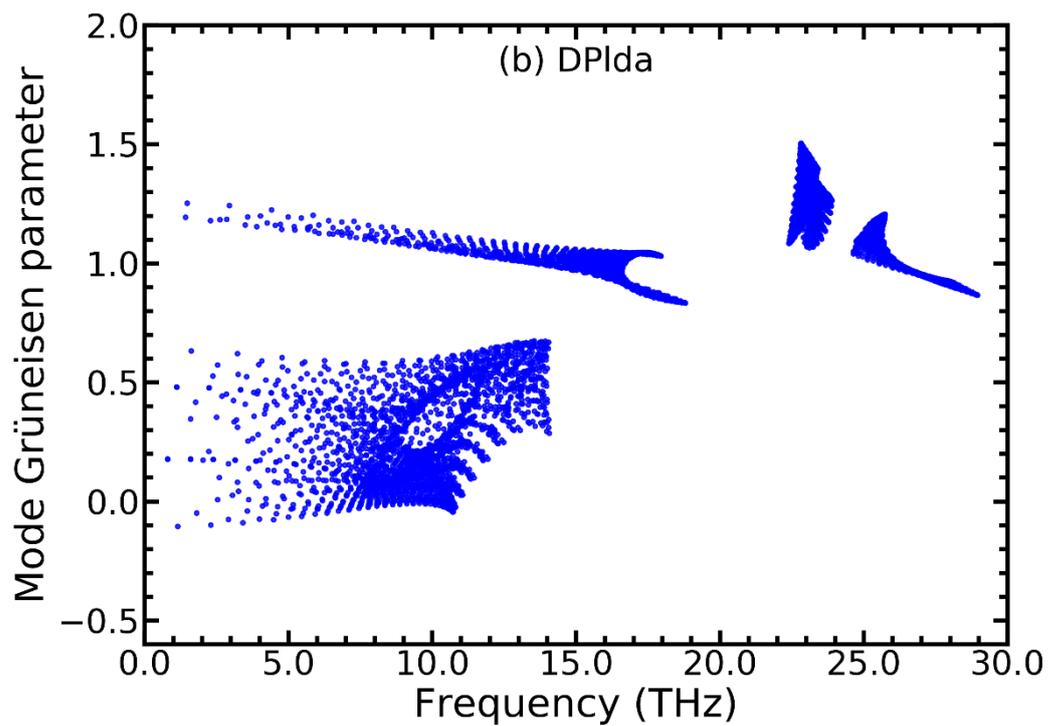

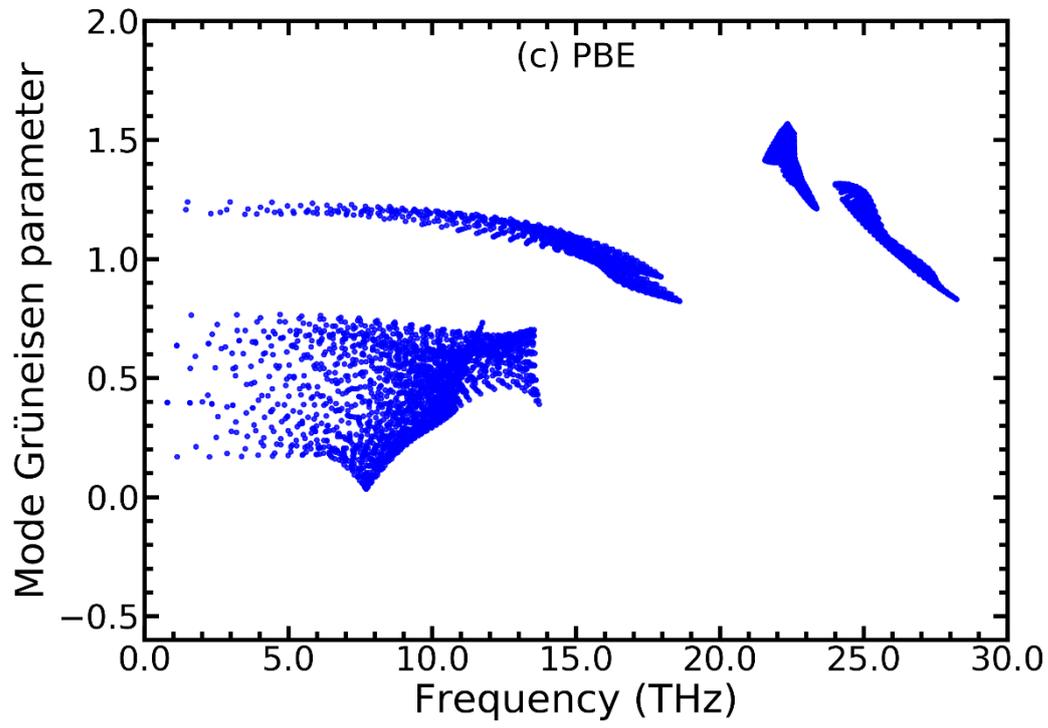

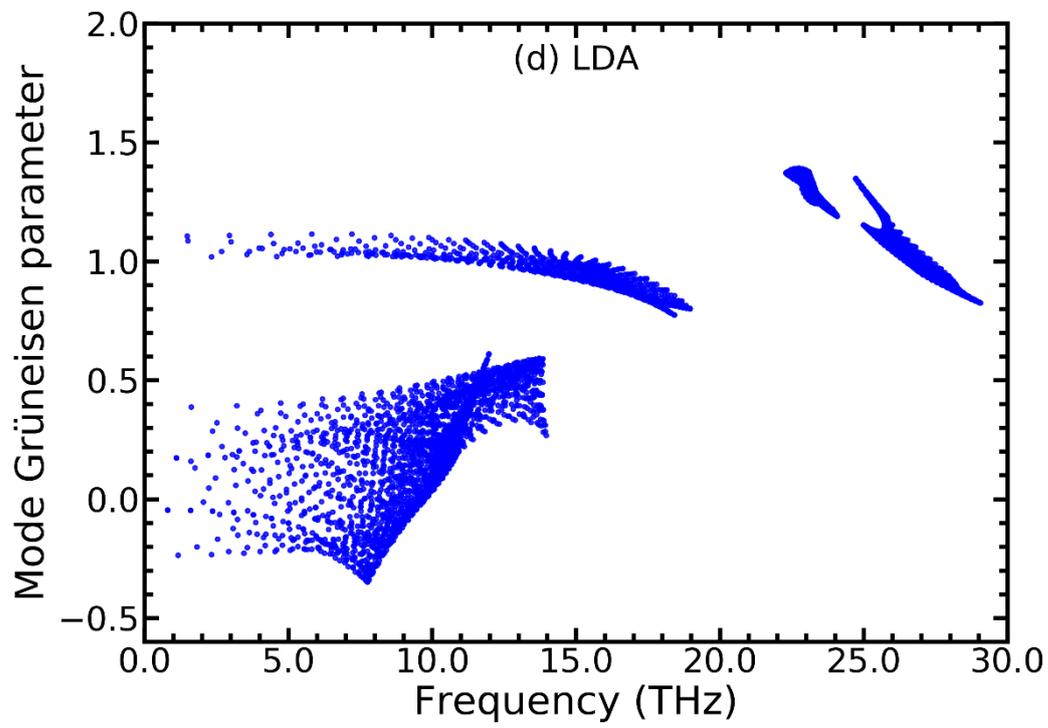

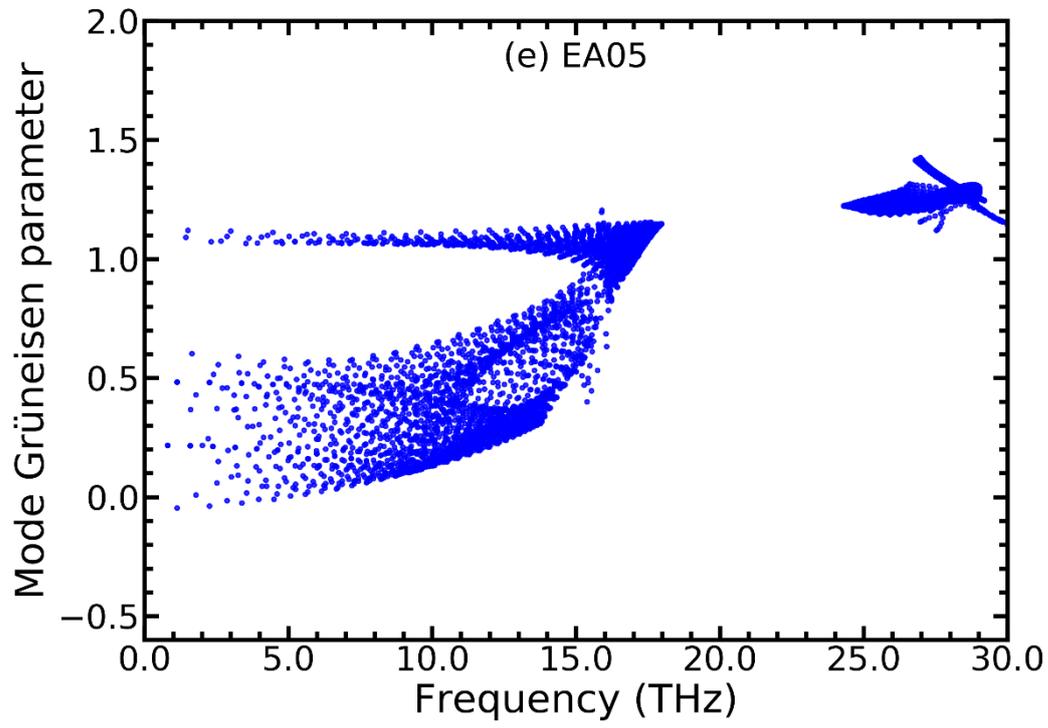

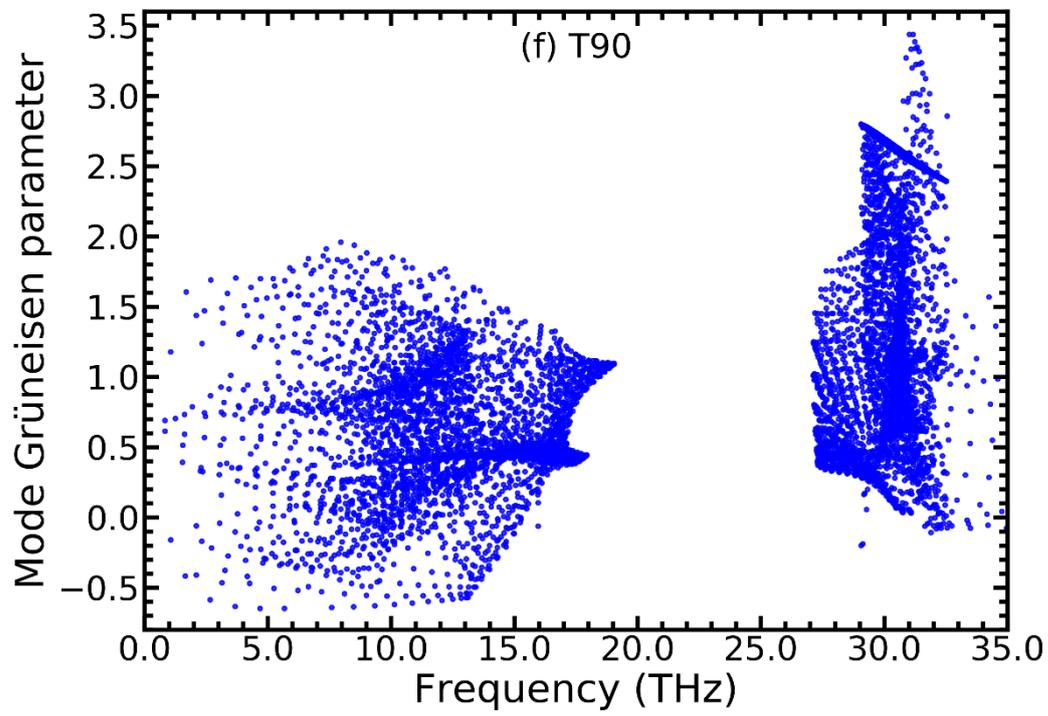